\documentclass[aps,a4paper,superscriptaddress,nofootinbib,twocolumn]{revtex4-1}
\usepackage{amsmath,amssymb,gensymb,graphicx,multirow,dcolumn,bm,latexsym,soul,nicefrac,acronym}
\usepackage[mathscr]{euscript}
\usepackage{appendix}
\usepackage[colorlinks,linkcolor=blue,citecolor=blue,urlcolor=blue ]{hyperref}
\usepackage{float}
\usepackage{soul}
\usepackage{verbatim}
\usepackage{color}
\usepackage{acronym}
\usepackage{subfigure}
\usepackage{appendix}
\usepackage{longtable}
\usepackage{footnote}
\usepackage{cancel}
\usepackage{ulem} % used solely for the purpose of enable strike-through lines. Should remove/comment before submission

\newcommand{\mnras}{Monthly Notices of the Royal Astronomical Society}

\newcommand{\be}{\begin{equation}}
\newcommand{\ee}{\end{equation}}
\newcommand{\bea}{\begin{eqnarray}}
\newcommand{\eea}{\end{eqnarray}}
\newcommand{\bw}{\begin{widetext}}
\newcommand{\ew}{\end{widetext}}
\newcommand{\nn}{\nonumber}

\newcommand{\eq}[1]{Eq.~(\ref{#1})}
\newcommand{\fig}[1]{Fig.~\ref{#1}}
\newcommand{\tab}[1]{Tab.~\ref}

\newcommand{\TRC}{MOE Key Laboratory of TianQin Mission, TianQin Research Center for Gravitational Physics $\&$  School of Physics and Astronomy, Frontiers Science Center for TianQin, CNSA Research Center for Gravitational Waves, Sun Yat-sen University (Zhuhai Campus), Zhuhai 519082, China}

\begin{document}
\title{Impact of combinations of time-delay interferometry channels on stochastic gravitational wave background detection}
\author{Zheng-Cheng Liang}
%\email{liangzhch7@mail.sysu.edu.cn}
\affiliation{\TRC}
%\affiliation{\HUST}
\author{Zhi-Yuan Li}
\affiliation{\TRC}
\author{Jun Cheng}
\affiliation{\TRC}
\author{En-Kun Li}
\affiliation{\TRC}
\author{Jian-dong Zhang}
%\email{zhangjd9@sysu.edu.cn}
\affiliation{\TRC}
\author{Yi-Ming Hu}
\email{huyiming@sysu.edu.cn}
\affiliation{\TRC}

\date{\today}

\begin{abstract}
The method of time delay interferometry (TDI) is proposed to cancel the laser noise in space-borne gravitational-wave detectors. 
Among all different TDI combinations, the most commonly used ones are the orthogonal channels A, E and T, where A and E are signal-sensitive and T is signal-insensitive. 
Meanwhile, for the detection of stochastic gravitational-wave background, one needs to introduce the overlap reduction function to characterize the correlation between channels. 
For the calculation of overlap reduction function, it is often convenient to work in the low-frequency approximation, and assuming the equal-arm Michelson channels. 
However, if one wishes to work on the overlap reduction function of $\rm A/E$ channels, then the low-frequency approximation fails. 
We derive the exact form of overlap reduction function for $\rm A/E$ channels. 
Based on the overlap reduction function, we calculate the sensitivity curves of TianQin, TianQin I+II and TianQin + LISA. 
We conclude that the detection sensitivity calculated with $\rm A/E$ channels is mostly consistent with that obtained from the equal-arm Michelson channels. 
\end{abstract}

\keywords{}

\pacs{04.25.dg, 04.40.Nr, 04.70.-s, 04.70.Bw}

%%%%%%%%%%%%%%%%%%%%%%%%%%%%%%%%%%%%%%%%%%%%%%%%%%%%%%%%%%%%%%%%
\maketitle
\acrodef{SGWB}{stochastic \ac{GW} background}
\acrodef{GW}{gravitational-wave}
\acrodef{CBC}{compact binary coalescence}
\acrodef{MBHB}{supermassive black hole binary}
\acrodef{SBBH}{stellar-mass binary black hole}
\acrodef{EMRI}{extreme-mass-ratio inspiral}
\acrodef{DWD}{double white dwarf}
\acrodef{BH}{black hole}
\acrodef{NS}{neutron star}
\acrodef{BNS}{binary neutron star}
\acrodef{LIGO}{Laser interferometry Gravitational-Wave Observatory}
\acrodef{TQ}{TianQin}
\acrodef{LISA}{Laser interferometry Space Antenna}
\acrodef{KAGRA}{Kamioka Gravitational Wave Detector}
\acrodef{ET}{Einstein telescope}
\acrodef{DECIGO}{DECi-hertz interferometry GravitationalWave Observatory}
\acrodef{CE}{Cosmic Explorer}
\acrodef{NANOGrav}{The North American Nanohertz Observatory for Gravitational Waves}
\acrodef{LHS}{left-hand side}
\acrodef{RHS}{right-hand side}
\acrodef{ORF}{overlap reduction function}
\acrodef{ASD}{amplitude spectral density}
\acrodef{PSD}{power spectral density}
\acrodef{SNR}{signal-to-noise ratio}
\acrodef{TDI}{time delay interferometry}
\acrodef{PIS}{peak-integrated sensitivity}
\acrodef{PLIS}{power-law integrated sensitivity}
\acrodef{GR}{general relativity}
\acrodef{PBH}{primordial black hole}
\acrodef{SSB}{solar system baryo}
\acrodef{PT}{phase transition}
\acrodef{SM}{Standard Model}
\acrodef{EWPT}{electroweak phase transition}
\acrodef{PTA}{Pulsar Timing Arrays}
\acrodef{RD}{radiation-dominated}
\acrodef{MD}{matter-dominated}
\acrodef{NG}{Nambu-Goto}
%%%%%%%%%%%%%%%%%%%%%%%%%%%%%%%%%%%%%%%%%%%%%%%%%%%%%%%%%%%%%%%%
%%%% ±êÌâÒ³œáÊø %%%%%%%%%%%%%%%%%%%%%%%%%%%%%%%%%%%%%%%%%%%%%%%%
%%%%%%%%%%%%%%%%%%%%%%%%%%%%%%%%%%%%%%%%%%%%%%%%%%%%%%%%%%%%%%%%
%%%% µÚÒ»œÚ¿ªÊŒ %%%%%%%%%%%%%%%%%%%%%%%%%%%%%%%%%%%%%%%%%%%%%%%%
%%%%%%%%%%%%%%%%%%%%%%%%%%%%%%%%%%%%%%%%%%%%%%%%%%%%%%%%%%%%%%%%

\section{Introduction}
%=============Definition and origin of SGWB==============%

A \ac{SGWB} is formed by the incoherent superposition of plenty of unresolved \acp{GW}~\cite{Maggiore:1999vm,Christensen:2018iqi,Romano:2019yrj}. 
Especially, \ac{SGWB} will become foreground when exceeding the detector noise level~\cite{Robson:2019,Huang:2020rjf}. 
The origin of \ac{SGWB} can be generally divided into astrophysics and cosmology~\cite{deAraujo:2000gw,Martinovic:2020hru}. 
The astrophysical-origin contains nearby objects, among which the Galactic \ac{DWD} is predicted to produce an anisotropic foreground~\cite{Timpano:2005gm,Nissanke:2012eh}. 
The cosmological-origin is related to the physical processes of the early Universe, 
and the cosmological \ac{SGWB} is generally considered to be highly isotropic unless there is a specific physical mechanism~\cite{Liu:2020mru}. 

%========Detection channel==========%
Currently, the laser interferometry is applied to detect \acp{GW}. 
The typical laser interferometer is equal-arm Michelson consisting of four laser links. 
The equal-arm Michelson has been employed for ground-based \ac{GW} detectors~\cite{LIGOScientific:2014pky,VIRGO:2014yos}, where the laser noise experiences the same delays in the links and hence cancels. 
Unlike ground-based \ac{GW} detectors, space-borne \ac{GW} detectors have unequal armlengths due to the movement of satellites~\cite{TianQin:2015yph,LISA:2017pwj}. 
Thus, it is hardly to construct an equal-arm Michelson for space-borne \ac{GW} detectors, and the laser noise will dominate the detector noise~\cite{Tinto:1995sd}. 
To cancel the laser noise in the unequal-arm detectors, Tinto et.al~\cite{Tinto:1999yr,1999ApJ...527..814A} proposed the \ac{TDI} technique adopting specific combinations of laser links. 
Among all possible \ac{TDI} combinations, the \ac{TDI} channel $\rm A$, $\rm E$ and $\rm T$ are the most commonly used, and they form the orthogonal channel group $\rm AET$. 
In the \ac{GW} detection, $\rm A/E$ channels are sensitive to \acp{GW}, while $\rm T$ channel works effectively for monitoring the detector noise. 

%========detection strategy==========%
Depending on whether more than one detector works at the same time, two different detection strategies are raised. 
One is cross-correlation method~\cite{Hellings:1983fr,1987MNRAS.227..933M,Christensen:1992wi,Flanagan:1993ix}, which is applied to the scenario that multiple detectors are employed to detect a common \ac{SGWB}. 
Under the assumption that the noise of two detectors is uncorrelated, one can distinguish the \ac{SGWB} signal from the detector noise by correlating the outputs from two detectors. 
The other one is null-channel method~\cite{Tinto:2001ii,Hogan:2001jn}, which is proposed for a single detector. 
%~\cite{Cutler:1997ta,Adams:2010vc}
In this method, detector noise is monitored by the null channel, 
so that the \ac{SGWB} signal can be extracted by auto-correlating the output from the detector. 

%========ORF=========%
When employing the above two methods to detect \ac{SGWB}, the key is the auto- or cross-correlation of the detector channel. 
To indicate the correlation, one needs to introduce the frequency-dependent correlation coefficient of \ac{SGWB} signals, i.e., \ac{ORF}~\cite{Christensen:1992wi,Flanagan:1993ix}. 
\ac{ORF} is determined by: (i) detector orientation; (ii) detector separation, which denotes the distance between the \ac{GW} measurement locations of the detector channel. 
For the equal-arm Michelson consisting of one laser interference site, the detector separation can be directly defined as the distance between the interference site of both equal-arm Michelsons. 
However, $\rm A/E$ channels contain multiple laser interference sites due to their construction~\cite{Vallisneri:2012np}, it is nontrivial to define the detector separation. 
%There are several ways to calculate the \ac{ORF} without defining the detector separation for A/E channels:  
Then by simplifying $\rm A/E$ channels to two orthogonal equal-arm Michelsons under the low-frequency approximation, Seto et al. calculated the \ac{ORF} for multiple triangular detectors~\cite{Seto:2020mfd}. 
Splitting the \ac{TDI} channel into component links, Hu et al. derived the \ac{ORF} for any \ac{TDI} channel without low-frequency approximation, based on which they calculated the \ac{ORF} of $\rm A/E$ channels for TianQin + LISA~\cite{Hu:2022byd}. 

%========SGWB detection assessment=========%
\ac{ORF} only involves the correlation between the detector channel, for which if one wants to demonstrate the detection capability, the detector noise level also needs to be considered. 
In general, detector noise level is characterized by noise \ac{PSD}. 
Then based on the \ac{ORF} and noise \ac{PSD}, the sky-average sensitivity curve is introduced~\cite{Larson:1999we}. 
Furthermore, the improvement in sensitivity that comes from the accumulation of correlation time and integrating over frequency also should be illustrated for \ac{SGWB} detection~\cite{Allen:1997ad}. 
Therefore, two more appropriate sensitivity curves, namely \ac{PLIS} curve~\cite{Thrane:2013oya} and \ac{PIS} curve~\cite{Schmitz:2020syl} are proposed. 

%===========What we have done========%
This paper aims to investigate the impacts of $\rm A/E$ channels on \ac{SGWB} detection. 
First, we select a reference interference site for the detector channel, then the distance between the reference interference sites can be defined as the detector separation. 
Following this rule, one can calculate the \ac{ORF} for any channel.  
Next, for TianQin, TianQin I+II and TianQin + LISA, we calculate the \acp{ORF} of the equal-arm Michelson and $\rm A/E$ channels without low-frequency approximation, respectively. 
Based on the \acp{ORF}, we further draw the corresponding \ac{PLIS} and \ac{PIS} curves. 
We find that the sensitivity curve of $\rm A/E$ channels is basically consistent with that of the equal-arm Michelson. 

%========================outline=========================%
The outline of this paper is as follows. 
In Sec.~\ref{sec:spectrum}, we introduce the strength of \ac{SGWB}. 
The channel response to \ac{SGWB} is discussed in Sec.~\ref{sec:response}. 
The \ac{ORF} is derived in Sec.~\ref{sec:ORF}. 
Sec.~\ref{sec:method} and~\ref{sec:sensitivity} are applied to assess \ac{SGWB} detection. 
Our conclusions are discussed in Sec.~\ref{sec:conclusion}. 
In addition, we show the response to \ac{GW} for equal-arm Michelson in~Appendix~\ref{appen:channel_response}. 
The discussion on the \ac{ORF} and noise level for channels are in~Appendix~\ref{appen:ORF_Michelson} and~Appendix~\ref{appen:noise}, respectively. 
The derivation for cross-correlation and null-channel methods is shown in~Appendix~\ref{appen:Q_W}.

\section{Strength of stochastic gravitational-wave background}\label{sec:spectrum} 
In our work, we focus on stationary, unpolarized and Gaussian \ac{SGWB}. Then the statistical properties of the \ac{SGWB} can be characterized by the dimensionless energy spectral density $\Omega_{\rm gw}(f)$, which is normalized by the critical energy density $\rho_{\rm c}$~\cite{Allen:1996vm}: 
\be
\label{eq:omega_gw}
\Omega_{\rm gw}(f)=\frac{1}{\rho_{\rm c}}\frac{{\rm d}\rho_{\rm gw}}{{\rm d}(\ln{f})},
\ee
where $\rho_{\rm c}\equiv3H_{0}^{2}c^{2}/(8\pi G)$ with the light speed $c$, the gravitational constant $G$, and the Hubble constant $H_{0}$. 
%Because different measurement methods get different values of $H_{0}$~\cite{Aghanim:2018eyx,Riess:2021jrx}, 
In general, we can adopt $h_{0}^{2}\Omega_{\rm gw}$ rather than $\Omega_{\rm gw}$ to remove the measurement uncertainty of $H_{0}$, where $H_{0}=h_{0}\times100\,{\rm km\,s^{-1}\,Mpc^{-1}}$. 
${\rm d}\rho_{\rm gw}$ denotes the \ac{GW} energy density $\rho_{\rm gw}$ stored in the frequency segment ${\rm d}f$, and in terms of the transverse-traceless metric perturbation $h_{ab}(t,\vec{x})$, \ac{GW} energy density can be defined as~\cite{Misner:1974qy,Creighton:2011zz}
\be
\label{eq:rho_gw}
\rho_{\rm gw}=\frac{c^2}{32\pi G}
\langle\dot{h}_{ab}(t,\vec{x})\dot{h}^{ab}(t,\vec{x})\rangle,
\ee
where $\langle\rangle$ indicates averaging over several wavelengths or periods of the \ac{GW}. 

Since the \ac{SGWB} is a collection of a large number of \acp{GW}, the transverse-traceless metric perturbation of the \ac{SGWB} can be expanded as a superposition of the \ac{GW} with wave vector $\hat{k}$:
\bea
\label{eq:h_ab}
h_{ab}(t,\vec{x})&=&\int_{-\infty}^{\infty}{\rm d}f\int_{S^{2}}{\rm d}\hat{\Omega}_{\hat{k}}\,
\widetilde{h}_{ab}(f,\hat{k})e^{{\rm i}2\pi f[t-\hat{k}\cdot\vec{x}(t)/c]},
\eea
where $h_{ab}(t,\vec{x})=h_{ab}(t-\hat{k}\cdot\vec{x}/c,\vec{0})$. 
In terms of polarization modes $P=+,\times$ and polarization
tensors $e^{P}_{ab}(\hat{k})$, the Fourier amplitude $\widetilde{h}_{ab}(f,\hat{k})$ can be expressed as: $\widetilde{h}_{ab}(f,\hat{k})=\sum_{P=+,\times}\widetilde{h}_{P}(f,\hat{k})e^{P}_{ab}(\hat{k})$. 
The amplitude $\widetilde{h}_{P}(f,\hat{k})$ is a random value with zero-mean, and the conjugate symmetry of Fourier transform holds: $\widetilde{h}_{P}(f,\hat{k})=\widetilde{h}^{*}_{P}(-f,\hat{k})$. 

Assuming \ac{SGWB} is stationary, the \ac{PSD} of \ac{SGWB} in propagation direction $\hat{k}$ can be defined by
\be
\label{eq:Ph}
\langle\widetilde{h}_{P}(f,\hat{k})\widetilde{h}^{*}_{P'}(f',\hat{k}')\rangle
=\frac{1}{4}\delta(f-f')\delta_{PP'}\delta^{2}(\hat{k}-\hat{k}')\mathscr{P}_{\rm h}(f,\hat{k}),
\ee
where the factor of $1/4$ agrees with the one-sided \ac{PSD} and the contribution of each polarization. 
The \ac{PSD} $\mathscr{P}_{\rm h}(f,\hat{k})$ is factorized by angular distribution $\mathcal{P}_{\rm h}(\hat{k})$ and spectral density $\bar{H}(f)$: $\mathscr{P}_{\rm h}(f,\hat{k})=\mathcal{P}_{\rm h}(\hat{k})\bar{H}(f)$. 
The spectral density can be further normalized by the reference frequency: $\bar{H}(f)=\bar{H}(f_{\rm ref})(f/f_{\rm ref})^{\epsilon}$, where the pow-law index $\epsilon$ depends on the origin of \ac{SGWB}~\cite{Thrane:2009fp}. 
For an anisotropic \ac{SGWB}, angular distribution $\mathcal{P}_{\rm h}(\hat{k})$ is usually decomposed into spherical harmonics~\cite{Allen:1996gp}:
\be
\mathcal{P}_{\rm h}(\hat{k})
=\sum_{l=0}^{\infty}\sum_{m=-l}^{l}\mathscr{P}_{lm}Y_{lm}(\hat{k}).
\ee
Furthermore, the sum of $\mathscr{P}_{\rm h}$ in all directions is the one-sided strain \ac{PSD} $S_{\rm h}(f)$ of \ac{SGWB}:
\bea
\label{eq:S2P}
S_{\rm h}(f)&=&\int_{S^{2}}{\rm d}\hat{\Omega}_{\hat{k}}\mathscr{P}_{\rm h}(f,\hat{k}),
\eea
and for an unpolarized \ac{SGWB}: 
\be
S^{+}_{\rm h}(f)=S^{\times}_{\rm h}(f)=\frac{1}{2}S_{\rm h}(f).
\ee
Then in terms of \eq{eq:h_ab}, \eq{eq:Ph} and \eq{eq:S2P}, the energy density of \ac{SGWB} can be calculated by
\bea
\label{rho2Sh}
\nn
\rho_{\rm gw}
&=&\frac{c^2}{32\pi G}
\int_{-\infty}^{\infty}{\rm d}f\int_{-\infty}^{\infty}{\rm d}f'
\int_{S^{2}}{\rm d}\hat{\Omega}_{\hat{k}}\int_{S'^{2}}{\rm d}
\Omega_{\hat{k}'}\\
\nn
&\quad&\times
4\pi^{2}\sum_{PP'}
\langle\widetilde{h}_{P}(f,\hat{k})\widetilde{h}^{*}_{P'}(f',\hat{k}')\rangle
e^{P}_{ab}(\hat{k})e_{P'}^{ab}(\hat{k}')\\
\nn
&\quad&\times
e^{{\rm i}2\pi[(f-f')t-(\hat{k}-\hat{k}')\cdot\vec{x}(t)/c]}\\
&=&\frac{\pi c^{2}}{4G}\int_{0}^{\infty}{\rm d}f\,f^{2}S_{\rm h}(f),
\eea

Combined with \eq{eq:omega_gw} and \eq{rho2Sh}, the energy spectral density $\Omega_{\rm gw}(f)$ can be converted to $S_{\rm h}(f)$ through
\be
\label{eq:Omega2Sh}
\Omega_{\rm gw}(f)
=\frac{2\pi^{2}}{3H_{0}^{2}}f^{3}S_{\rm h}(f).
\ee
Note that in some studies, $S_{\rm h}(f)$ only contains the contribution of one polarization, so the factor of 2 on the molecule of \eq{eq:Omega2Sh} turns to 4~\cite{Cornish:2001bb}.
%Both $\Omega_{\rm gw}(f)$ and $S_{\rm h}(f)$ are intrinsic to \ac{SGWB}, which does not rely on the detection process. 

\section{Channel response to stochastic gravitational-wave background}\label{sec:response}
In this section, we further discuss the response of channels to \ac{SGWB} on the detector. 
Generally speaking, the channel response to \ac{SGWB} always changes due to detector motion. 
However, it is reasonable to assume that the response is nearly unchanged within a sufficiently short period $[t_{0}-T/2,t_{0}+T/2]$. 
Then with the help of short-term Fourier transform, the \ac{SGWB} signal that is a convolution of the metric perturbation $h_{ab}(t,\vec{x})$ and the impulse response $\mathbb{D}^{ab}(t,\vec{x})$ can be expressed as~\cite{Romano:2016dpx}: 
\bea
\label{eq:ht_sgwb}
\nn
h(t,t_{0})&=&\mathbb{D}^{ab}[t,\vec{x}(t_{0})]*h_{ab}[t,\vec{x}(t_{0})]\\
\nn
&=&\sum_{P=+,\times}\int_{-\infty}^{\infty}{\rm d}f\int_{S^{2}}{\rm d}\hat{\Omega}_{\hat{k}}
F^{P}(f,\hat{k},t_{0})\widetilde{h}_{P}(f,\hat{k})\\
&\times&e^{{\rm i}2\pi f[t-\hat{k}\cdot\vec{x}(t_{0})/c]},
\eea
where $\vec{x}$ labels the location where \ac{GW} measurement occurs. 
The response function can be decomposed as $F^{P}(f,\hat{k},t_{0})=e^{P}_{ab}(\hat{k})F^{ab}(f,\hat{k},t_{0})$ with
\bea
\nn
F^{ab}(f,\hat{k},t_{0})&=&
\int_{t_{0}-T/2}^{t_{0}+T/2}{\rm d}\,\tau\int{\rm d}^{3}\vec{y}\,
\mathbb{D}^{ab}[\tau,\vec{y}(t_{0})]\\
&\quad&\times
e^{-{\rm i}2\pi f[\tau-\hat{k}\cdot\vec{y}(t_{0})/c]},
\eea
and the frequency domain signal is expressed as
\be
\label{eq:hf_sgwb}
\widetilde{h}(f,t_{0})
=\int_{S^{2}}\,{\rm{d}}\hat{\Omega}_{\hat{k}}
\sum_{P=+,\times}F^{P}(f,\hat{k},t_{0})\widetilde{h}_{P}(f,\hat{k})
e^{-{\rm i}2\pi f\hat{k}\cdot\vec{x}(t_{0})/c}.
\ee 

For ground-based \ac{GW} detectors~\cite{LIGOScientific:2014pky,VIRGO:2014yos}, the equal-arm Michelson can be employed to detect \acp{GW} because the armlength keeps unchanged. 
When it comes to space-borne detectors~\cite{LISA:2017pwj,TianQin:2015yph}, the armlength is difficult to be maintained and the \ac{TDI} channel is introduced to cancel the laser noise~\cite{Tinto:2001ii,Tinto:2002de,Hogan:2001jn}. 
However, to facilitate the discussion for the impacts of the \ac{TDI} channel A/E on \ac{SGWB} detection, we will construct both the equal-arm Michelson and $\rm A/E$ channels in the same regular triangle detector. 
\begin{figure}[t]
\centering
\includegraphics[height=7.5cm]{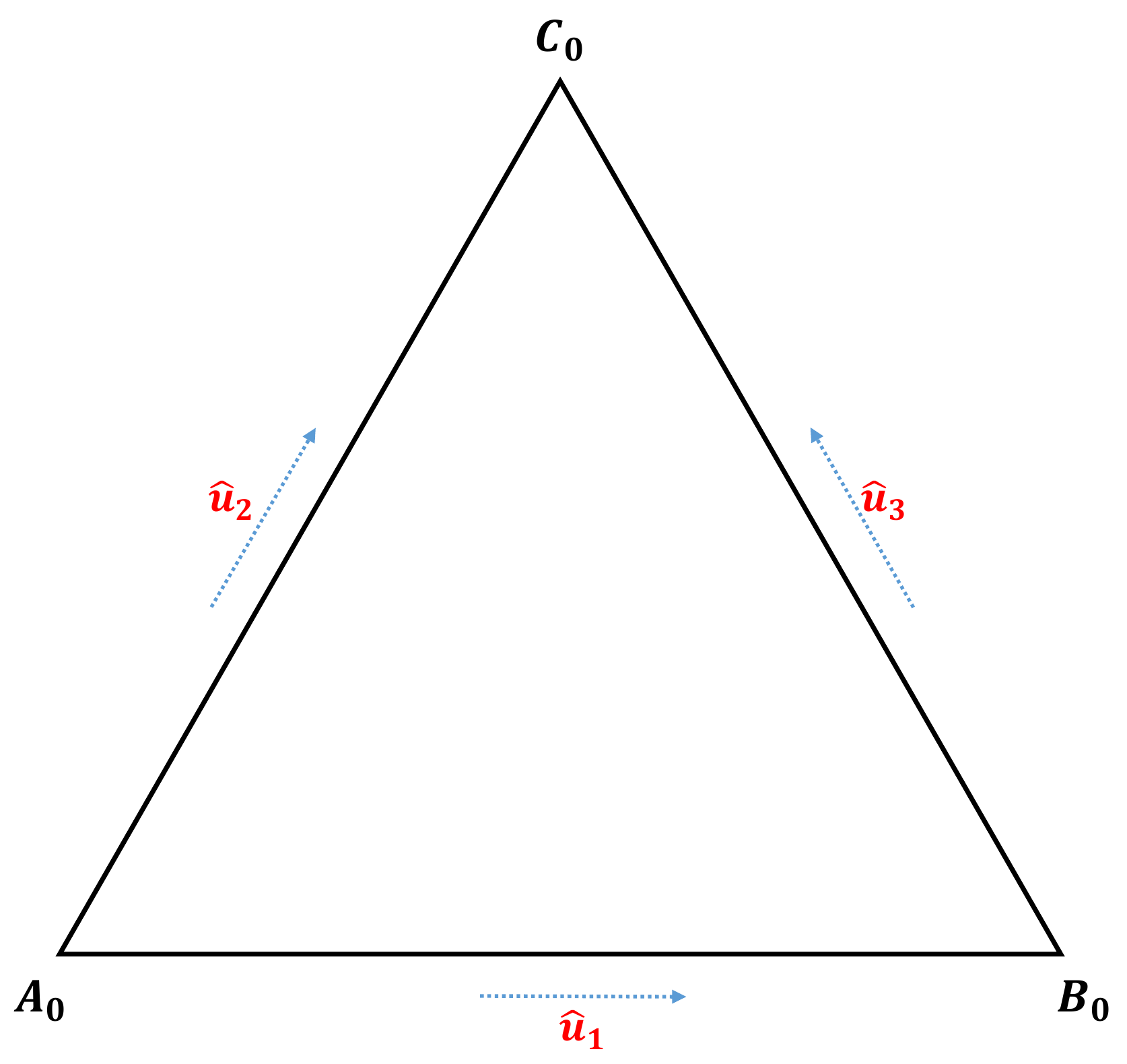}
\caption{Schematic diagram of a regular triangle detector.}
\label{fig:Triangle}
\end{figure}

As shown in~\fig{fig:Triangle}, we can construct an equal-arm Michelson by a vertex and two adjacent edges. 
Following this rule, the equal-arm Michelson channel group $\rm M_{1}M_{2}M_{3}$ can be constructed based on the corner satellite $A_{0}$, $B_{0}$ and $C_{0}$, respectively. 
In addition, by defining the round trip of the laser in one arm as two-way tracking, the equal-arm Michelson consists of two different two-way trackings. 
Then the response functions of channel group $\rm M_{1}M_{2}M_{3}$ can be written as
\bea
\label{eq:FM_abc}
\nn
F_{{\rm M}_{1}}^{P}(f,\hat{k},t_{0})
&=&F_{\rm II}^{P}[f,\hat{k},\hat{u}_{1}(t_{0})]-F_{\rm II}^{P}[f,\hat{k},\hat{u}_{2}(t_{0})],\\
\nn
F_{{\rm M}_{2}}^{P}(f,\hat{k},t_{0})
&=&F_{\rm II}^{P}[f,\hat{k},\hat{u}_{3}(t_{0})]-F_{\rm II}^{P}[f,\hat{k},-\hat{u}_{1}(t_{0})],\\
F_{{\rm M}_{3}}^{P}(f,\hat{k},t_{0})
\nn
&=&F_{\rm II}^{P}[f,\hat{k},-\hat{u}_{2}(t_{0})]-F_{\rm II}^{P}[f,\hat{k},-\hat{u}_{3}(t_{0})],\\
\eea
where $F_{\rm II}^{P}$ is the response function of the two-way tracking, and $\overrightarrow{A_{0}B_{0}}=L\hat{u}_{1}$, $\overrightarrow{A_{0}C_{0}}=L\hat{u}_{2}$, $\overrightarrow{B_{0}C_{0}}=L\hat{u}_{3}$ with the armlength $L$. 
More details are shown in Appendix~\ref{appen:channel_response}. 

In terms of the two responses of the same equal-arm Michelson with a time delay of $2L/c$, the response function of \ac{TDI} Michelson $\rm X$ can be obtained by
\bea
\nn
\label{eq:F_X}
F_{\rm X}^{P}(f,\hat{k},t_{0})
&=&F_{{\rm M}_{1}}^{P}(f,\hat{k},t_{0})-F_{{\rm M}_{1}}^{P}(f,\hat{k},t_{0}-2L/c)\\
&=&(1-e^{-i2f/f_{\ast}})F_{{\rm M}_{1}}^{P}(f,\hat{k},t_{0}).
\eea
Following the cyclic substitution, the response functions of $\rm Y$ and $\rm Z$ are
\bea
\label{eq:F_YZ}
\nn
F_{\rm Y}^{P}(f,\hat{k},t_{0})
&=&(1-e^{-i2f/f_{\ast}})F_{{\rm M}_{2}}^{P}(f,\hat{k},t_{0}),\\
F_{\rm Z}^{P}(f,\hat{k},t_{0})
&=&(1-e^{-i2f/f_{\ast}})F_{{\rm M}_{3}}^{P}(f,\hat{k},t_{0}).
\eea

Furthermore, the orthogonal channel group $\rm AET$ can be formed by channel group $\rm XYZ$~\cite{Vallisneri:2012np,Tinto:2014lxa,Cheng:2022vct}:
\bea
\label{eq:channel_AE}
\nn
{\rm A}&=&\frac{1}{\sqrt{2}}({\rm Z}-{\rm X}),\\
\nn
{\rm E}&=&\frac{1}{\sqrt{6}}({\rm X}-2{\rm Y}+{\rm Z}),\\
{\rm T}&=&\frac{1}{\sqrt{3}}({\rm X}+{\rm Y}+{\rm Z}).
\eea
Note that, the correlation between channel $\rm X$, $\rm Y$ and $\rm Z$ is not taken into account in \eq{eq:channel_AE}. 
Considering this correlation, Adams et.al~\cite{Adams:2010vc,Adams:2013qma} provide a new construction for channel group $\rm AET$. 
In this work, we still adopt \eq{eq:channel_AE} which is more widely used.

Through the above construction, there will be multiple laser interference sites in channel $\rm A$, $\rm E$ and $\rm T$. 
In order to derive the response functions of channel group $\rm AET$, one needs to calculate the corresponding \ac{SGWB} signals based on \eq{eq:hf_sgwb}:
\bw
\bea
\label{eq:hf_aet}
\nn
\widetilde{h}_{\rm A}(f,t_{0})
&=&\frac{1}{\sqrt{2}}\big(\widetilde{h}_{\rm Z}(f,t_{0})-\widetilde{h}_{\rm X}(f,t_{0})\big)\\
\nn
&=&\sum_{P=+,\times}\int_{S^{2}}\,{\rm{d}}\hat{\Omega}_{\hat{k}}
\,\frac{1}{\sqrt{2}}\big[F_{\rm Z}^{P}(f,\hat{k},t_{0})
e^{-{\rm i}2\pi f\hat{k}\cdot\overrightarrow{OC_{0}}(t_{0})/c}
-F_{\rm X}^{P}(f,\hat{k},t_{0})
e^{-{\rm i}2\pi f\hat{k}\cdot\overrightarrow{OA_{0}}(t_{0})/c}\big]
\widetilde{h}_{P}(f,\hat{k})\\
\nn
&=&\sum_{P=+,\times}\int_{S^{2}}\,{\rm{d}}\hat{\Omega}_{\hat{k}}
F_{\rm A}^{P}(f,\hat{k},t_{0})
\widetilde{h}_{P}(f,\hat{k})e^{-{\rm i}2\pi f\hat{k}\cdot\overrightarrow{OA_{0}}(t_{0})/c},\\
\nn
\widetilde{h}_{\rm E}(f,t_{0})
&=&\frac{1}{\sqrt{6}}\big(\widetilde{h}_{\rm X}(f,t_{0})-2\widetilde{h}_{\rm Y}(f,t_{0})+
\widetilde{h}_{\rm Z}(f,t_{0})\big)\\
\nn
&=&\sum_{P=+,\times}\int_{S^{2}}\,{\rm{d}}\hat{\Omega}_{\hat{k}}
F_{\rm E}^{P}(f,\hat{k},t_{0})
\widetilde{h}_{P}(f,\hat{k})e^{-{\rm i}2\pi f\hat{k}\cdot\overrightarrow{OA_{0}}(t_{0})/c},\\
\nn
\widetilde{h}_{\rm T}(f,t_{0})
&=&\frac{1}{\sqrt{3}}\big(\widetilde{h}_{\rm X}(f,t_{0})+\widetilde{h}_{\rm Y}(f,t_{0})+
\widetilde{h}_{\rm Z}(f,t_{0})\big)\\
&=&\sum_{P=+,\times}\int_{S^{2}}\,{\rm{d}}\hat{\Omega}_{\hat{k}}
F_{\rm T}^{P}(f,\hat{k},t_{0})
\widetilde{h}_{P}(f,\hat{k})e^{-{\rm i}2\pi f\hat{k}\cdot\overrightarrow{OA_{0}}(t_{0})/c},
\eea
where the vertex $A_{0}$ is selected to the reference interference site. 
Then the response functions are
\bea
\label{eq:F_aet}
\nn
F_{\rm A}^{P}(f,\hat{k},t_{0})&=&\frac{1}{\sqrt{2}}\big[F_{\rm Z}^{P}(f,\hat{k},t_{0})
e^{-{\rm i}2\pi f\hat{k}\cdot\overrightarrow{A_{0}C_{0}}(t_{0})/c}-F_{\rm X}^{P}(f,\hat{k},t_{0})\big],\\
\nn
F_{\rm E}^{P}(f,\hat{k},t_{0})&=&\frac{1}{\sqrt{6}}
\big[F_{\rm X}^{P}(f,\hat{k},t_{0})-2F_{\rm Y}^{P}(f,\hat{k},t_{0})
e^{-{\rm i}2\pi f\hat{k}\cdot\overrightarrow{A_{0}B_{0}}(t_{0})/c}
+F_{\rm Z}^{P}(f,\hat{k},t_{0})e^{-{\rm i}2\pi f\hat{k}\cdot\overrightarrow{A_{0}C_{0}}(t_{0})/c}\big],\\
F_{\rm T}^{P}(f,\hat{k},t_{0})&=&\frac{1}{\sqrt{3}}
\big[F_{\rm X}^{P}(f,\hat{k},t_{0})+F_{\rm Y}^{P}(f,\hat{k},t_{0})
e^{-{\rm i}2\pi f\hat{k}\cdot\overrightarrow{A_{0}B_{0}}(t_{0})/c}
+F_{\rm Z}^{P}(f,\hat{k},t_{0})
e^{-{\rm i}2\pi f\hat{k}\cdot\overrightarrow{A_{0}C_{0}}(t_{0})/c}\big].
\eea
Similarly, one can also construct channel group $\rm A'E'T'$ for another regular triangle detector with the response functions:
\bea
\label{eq:F_aet2}
\nn
F_{\rm A'}^{P}(f,\hat{k},t_{0})&=&\frac{1}{\sqrt{2}}\big[F_{\rm Z'}^{P}(f,\hat{k},t_{0})
e^{-{\rm i}2\pi f\hat{k}\cdot\overrightarrow{A_{0}'C_{0}'}(t_{0})/c}-
F_{\rm X'}^{P}(f,\hat{k},t_{0})\big],\\
\nn
F_{\rm E'}^{P}(f,\hat{k},t_{0})&=&\frac{1}{\sqrt{6}}
\big[F_{\rm X'}^{P}(f,\hat{k},t_{0})-
2F_{\rm Y'}^{P}(f,\hat{k},t_{0})
e^{-{\rm i}2\pi f\hat{k}\cdot\overrightarrow{A_{0}'B_{0}'}(t_{0})/c}
+F_{\rm Z'}^{P}(f,\hat{k},t_{0})
e^{-{\rm i}2\pi f\hat{k}\cdot\overrightarrow{A_{0}'C_{0}'}(t_{0})/c}\big],\\
F_{\rm T'}^{P}(f,\hat{k},t_{0})&=&\frac{1}{\sqrt{3}}
\big[F_{\rm X'}^{P}(f,\hat{k},t_{0})+F_{\rm Y'}^{P}(f,\hat{k},t_{0})
e^{-{\rm i}2\pi f\hat{k}\cdot\overrightarrow{A_{0}'B_{0}'}(t_{0})/c}
+F_{\rm Z'}^{P}(f,\hat{k},t_{0})
e^{-{\rm i}2\pi f\hat{k}\cdot\overrightarrow{A_{0}'C_{0}'}(t_{0})/c}\big],
\eea
\ew
where $A_{0}'$, $B_{0}'$, $C_{0}'$ are the vertexes, and $A_{0}'$ is the reference interference site. 

For the same detector, there is no correlation between the responses of $\rm A$, $\rm E$, and $\rm T$ to \ac{SGWB}, and $\rm T$ is the null channel where the response to \ac{SGWB} is highly suppressed. 
More details are shown in~Appendix~\ref{appen:ORF_Michelson}. 

\section{Overlap reduction function}\label{sec:ORF}
In order to characterize the statistical properties of \ac{SGWB} signal, one can introduce the correlation \ac{PSD}\footnote{The auto-correlation \ac{PSD} $P_{{\rm h}_{I}}$ is a real function.}:
\be
\label{eq:hIhJ}
\langle\widetilde{h}_{I}(f,t_{0})\widetilde{h}_{J}^{*}(f',t_{0})\rangle
=\frac{1}{2}\delta(f-f')P_{{\rm h}_{IJ}}(f,t_{0}),
\ee
where 
\be
P_{{\rm h}_{IJ}}(f,t_{0})=\Upsilon_{IJ}(f,t_{0})S_{\rm h}(f). 
\ee
Then we can connect the \acp{PSD} of \ac{SGWB} and \ac{SGWB} signal through the universal \ac{ORF} $\Upsilon_{IJ}$, which is independent of \ac{SGWB} spectral density $\bar{H}(f)$:
\be
\Upsilon_{IJ}(f,t_{0})
=\frac{\int_{S^{2}}{\rm d}\hat{\Omega}_{\hat{k}}\mathcal{Y}_{IJ}(f,\hat{k},t_{0})\mathcal{P}_{\rm h}(\hat{k})}{\int_{S^{2}}{\rm d}\hat{\Omega}_{\hat{k}}
\mathcal{P}_{\rm h}(\hat{k})}.
\ee
The geometric factor specifies the correlation between the responses of channel $I$ and $J$ to \ac{SGWB}:
\bea
\nn
\mathcal{Y}_{IJ}(f,\hat{k},t_{0})&=&\frac{1}{2}\sum_{P=+,\times}
F^{P}_{I}(f,\hat{k},t_{0})F^{P*}_{J}(f,\hat{k},t_{0})\\
&\quad&\times
e^{-{\rm i}2\pi f\hat{k}\cdot[\vec{x}_{I}(t_{0})-\vec{x}_{J}(t_{0})]/c},
\eea
where $\vec{x}_{I,J}$ denotes the laser interference sites of channel $I$ and $J$.  

For an isotropic \ac{SGWB}, the universal \ac{ORF} turns to the classical \ac{ORF}: 
\be
\label{eq:Gamma_IJ}
\Gamma_{IJ}(f,t_{0})=
\frac{1}{4\pi}\int_{S^{2}}{\rm d}\hat{\Omega}_{\hat{k}}
\mathcal{Y}_{IJ}(f,\hat{k},t_{0}).
\ee
For the case of the channel $\rm A$ and $\rm A'$ mentioned above, the \ac{ORF}
\bea
\label{eq:Gamma_AA'_rw}
\nn
\Gamma_{\rm AA'}(f,t_{0})
&=&\frac{1}{8\pi}\sum_{P=+,\times}\int_{S^{2}}{\rm d}\hat{\Omega}_{\hat{k}}
F_{\rm A}^{P}(f,\hat{k},t_{0})F_{\rm A'}^{P*}(f,\hat{k},t_{0})\\
&\quad&\times e^{-{\rm i}2\pi f\hat{k}\cdot\overrightarrow{A_{0}'A_{0}}(t_{0})/c},
\eea
where ${A_{0}'A_{0}}$ denotes the detector separation. 
Especially, the \ac{ORF} of one channel reduces to the transfer function $\mathcal{R}(f)$. 
\begin{figure}[t]
\centering
\includegraphics[height=6cm]{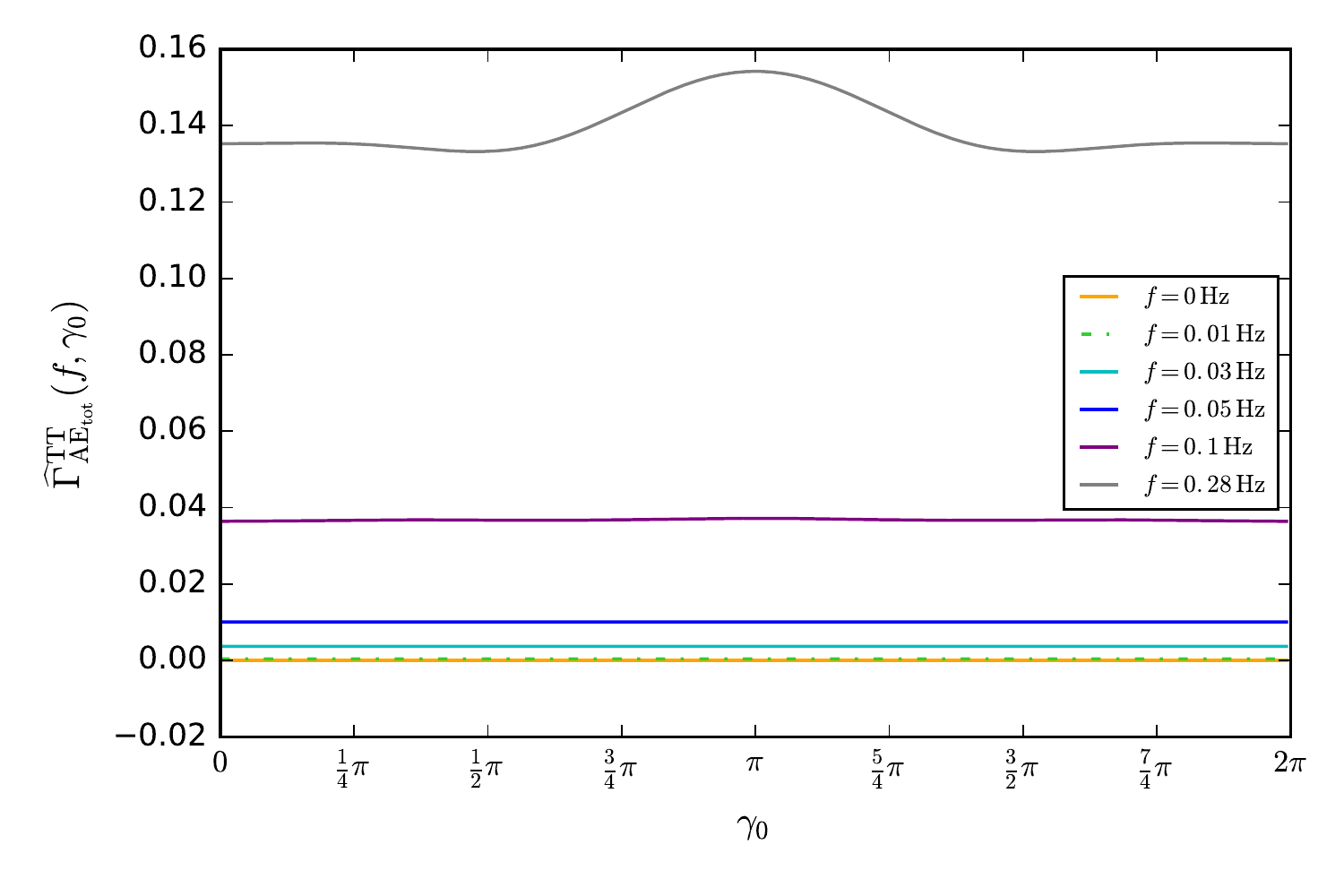}
\caption{Total \ac{ORF} of TianQin I+II for different initial angle difference $\gamma_{0}$.}
\label{fig:ORF_beta_2TQ}
\end{figure}
In the situation that \ac{ORF} changes over time, we need to further define the time-averaged \ac{ORF} for the total correlation time $T_{\rm tot}$~\cite{Liang:2021bde}:
\be
\bar{\Gamma}_{IJ}(f)=
\sqrt{\frac{1}{T_{\rm tot}}\int_{0}^{T_{\rm tot}}{\rm d}t\,|\hat{\Gamma}_{IJ}(f,t_{0})|^{2}}.
\ee

For TianQin I+II and TianQin + LISA, one can construct four pairs of channels for cross-correlation detection, then the {\it total} \ac{ORF} of TianQin I+II and TianQin + LISA is defined as~\cite{Seto:2020mfd}
\be
\label{eq:gamma_total}
\Gamma_{IJ_{\rm tot}}(f)=
\sqrt{\sum_{I,J}\big|\bar{\Gamma}_{IJ}(f)\big|^{2}},
\ee
where $I$ and $J$ label a pair of orthogonal equal-arm Michelsons or $\rm A/E$ channels, i.e., $\Gamma_{\rm{MM}_{\rm tot}}$ or $\Gamma_{\rm{AE}_{\rm tot}}$. Next, we will demonstrate the calculation of the \ac{ORF} for TianQin I+II and TianQin + LISA, respectively. 
And as a shorthand label, we might employ TQ for TianQin, TT for TianQin I+II, TL for TianQin + LISA in the figures and equations.

For TianQin I+II, since the orbital planes of TianQin and TianQin II will be perpendicular to each other all the time, the total \ac{ORF} will not change over time under the low-frequency approximation~\cite{Seto:2020mfd}. 
However, if low-frequency approximation fails, the total \ac{ORF} will be affected by the launch times of two detectors, i.e., the initial angles of TianQin and TianQin II~\cite{Liang:2021bde}. 
We denote the initial angular difference between TianQin and TianQin II with $\gamma_{0}$, and show the total \ac{ORF} of $\rm A/E$ channels within one orbital period of TianQin in \fig{fig:ORF_beta_2TQ}. 
We find that the optimal \ac{ORF} will be obtained when $\gamma_{0}=(2n+1)\pi$ with $n=0,1,2...$, i.e., the launch time difference between TianQin and TianQin II should be set to the semi-integer multiple of the orbital period. 
However, our previous work concluded that for the equal-arm Michelson, the launch time difference should be an integer multiple of the orbital period\footnote{It is shown in Fig.~4 of Ref.~\cite{Liang:2021bde}.}. 
Note that, there must be an overlap in the operation periods of TianQin and TianQin II, or the correlation time will drop to 0. 
Because the nominal working mode of TianQin is set to ``three months on + three months off", one needs to extend the operating time so that TianQin and TianQin II will run simultaneously for sufficient time. 
\begin{figure}[t]
\centering
\includegraphics[height=8.5cm]{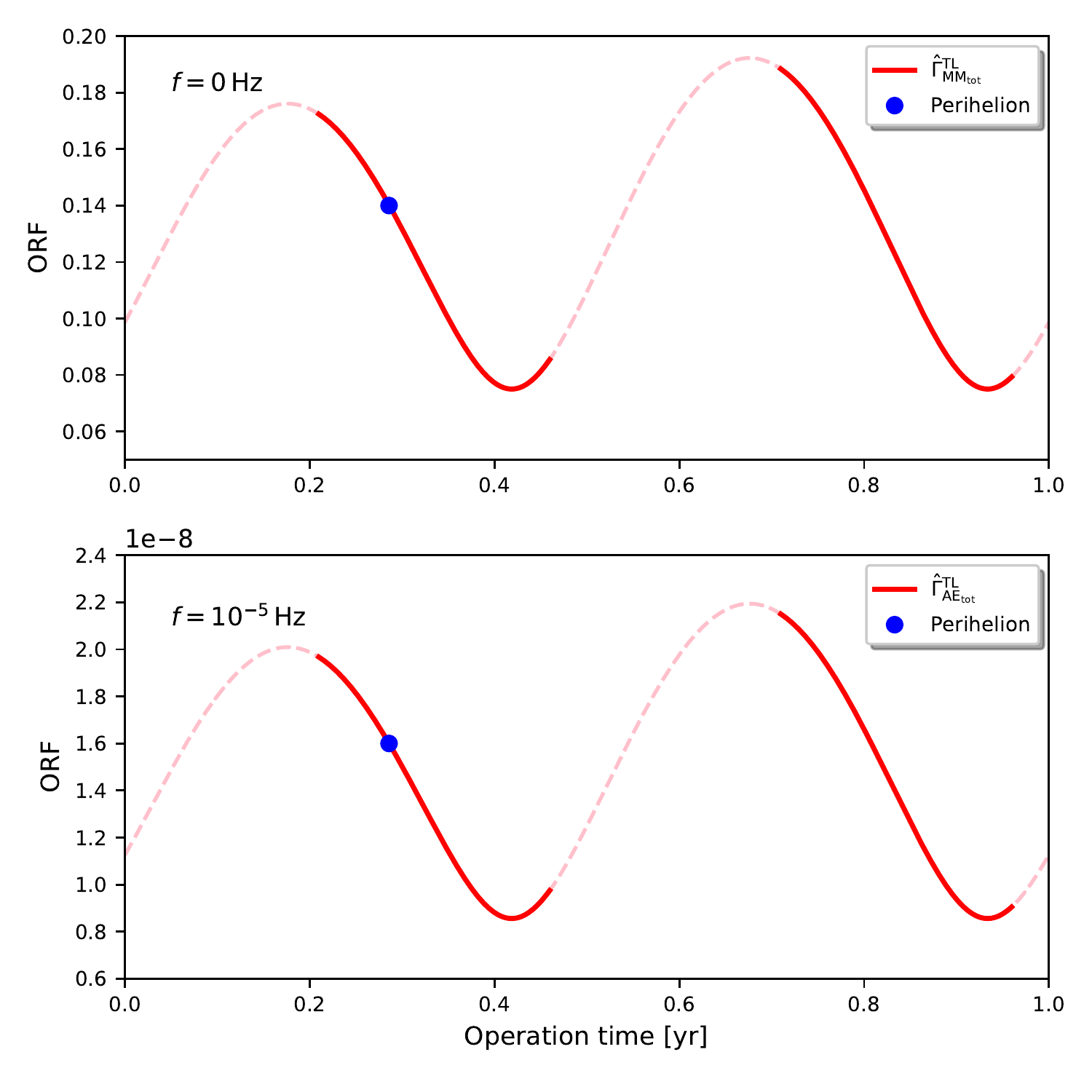}
\caption{\acp{ORF} of the equal-arm Michelson (top panel) and $\rm A/E$ channels (bottom panel) for TianQin + LISA.}
\label{fig:ORF_TL_low_f}
\end{figure}

As for TianQin + LISA, the angle between the orbital planes of TianQin and LISA will change periodically in a cycle of about one year~\cite{Liang:2021bde}. 
Besides, under the premise of LISA operating throughout the year, the TianQin + LISA configuration can perform cross-correlation detection only when TianQin is running. 
In terms of the nominal working mode of TianQin, we show the \ac{ORF} of TianQin + LISA throughout one year with low-frequency approximation in~\fig{fig:ORF_TL_low_f}. 
The top and bottom panels are the results of the equal-arm Michelson and $\rm A/E$ channels, respectively,  where the dashed part indicates that TianQin is off duty. 
Meanwhile, we mark the perihelion with blue dot\footnote{For the equal-arm Michelson, one can set $f=\rm 0\,\,Hz$ to get the \ac{ORF} under the low-frequency approximation; but for $\rm A/E$ channels, the \ac{ORF} drops to 0 when $f=0\,\,{\rm Hz}$, so we set $f=10^{-5}\,\,{\rm Hz}$.}. 
The primary and secondary peaks of both \acp{ORF} occur at $t=0.176,0.676\,\,{\rm yr}$, on which TianQin is off. 
Furthermore, since the orbital period of LISA is about 100 times that of TianQin, regardless of the initial angle of TianQin, the total \ac{ORF} of TianQin + LISA will not change after one year of cross-correlation detection. 

\begin{figure}[t]
\centering
\includegraphics[height=6cm]{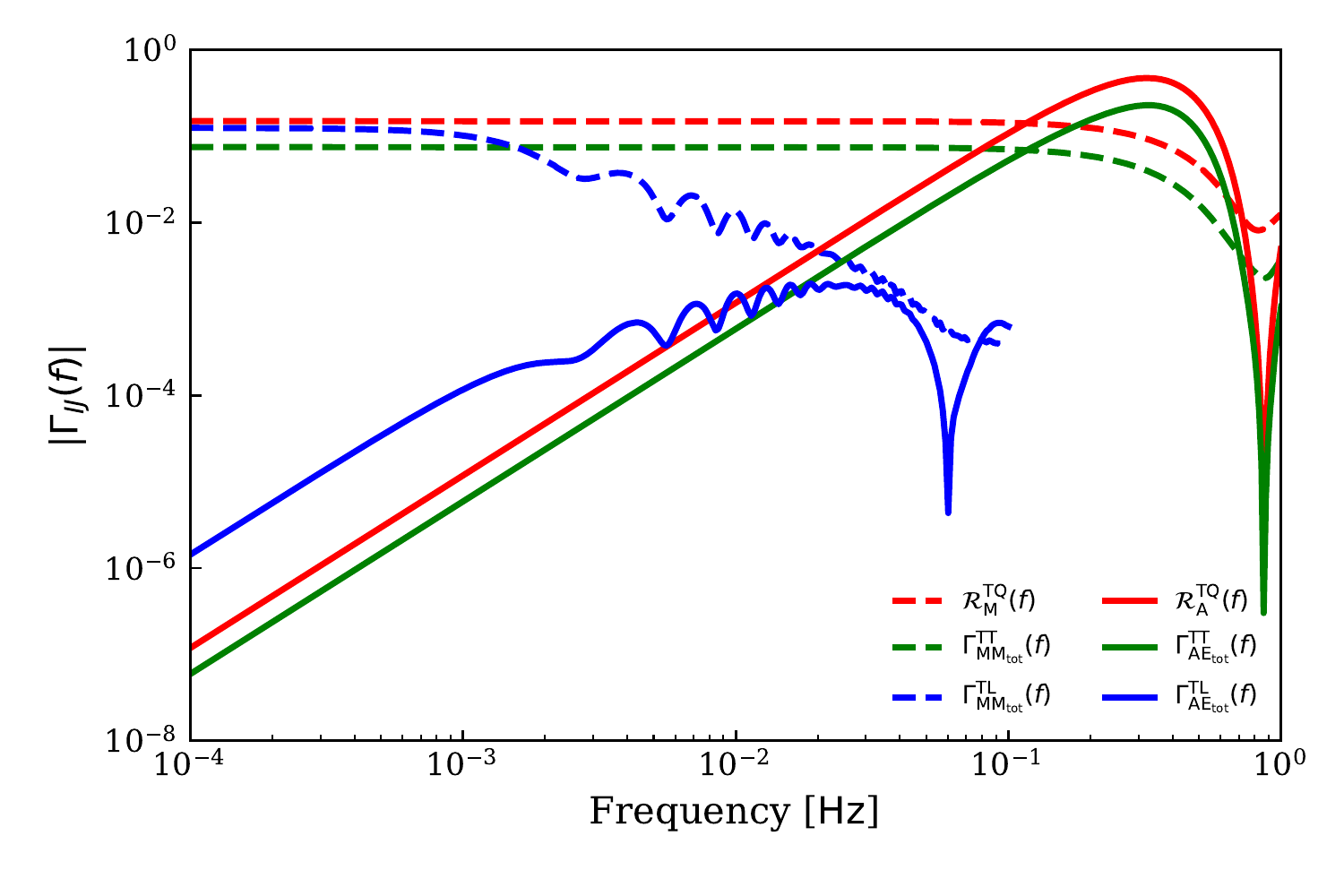}
\caption{Transfer function and \ac{ORF} for different detector configurations. Red, green and blue lines match TianQin, TianQin I+II and TianQin + LISA, while dashed and solid lines denote the results of the equal-arm Michelson and $\rm A/E$ channels, respectively. Due to the limitation of calculation accuracy, the \ac{ORF} for TianQin + LISA is truncated to 0.1 Hz.}
\label{fig:ORF_3config}
\end{figure}

Based on the above analysis, the total \acp{ORF} of TianQin I+II and TianQin + LISA are shown in~\fig{fig:ORF_3config}, where the transfer function of TianQin is also involved for comparison. 
Similar to the transfer function, the \ac{ORF} of equal-arm Michelson remains constant with low-frequency approximation, while for $\rm A/E$ channels it is proportional to $f^{2}$. 

%===============The detection \ac of SGWB===============%
\section{Detection method}\label{sec:method}
The output $s(t)$ of detector channel mainly contains \ac{SGWB} signal $h(t)$ and channel noise $n(t)$. 
Unless the \ac{SGWB} signal is much larger in magnitude than the channel noise, it is almost impossible to separate the signal from the noise through a single-channel measurement~\cite{Allen:1997ad}. 
The general strategies are cross-correlating the outputs from different noise-independent channels~\cite{Hellings:1983fr,Christensen:1992wi,Flanagan:1993ix} and auto-correlating the output from one channel under the noise monitoring of the null channel~\cite{Robinson:2008fb,Romano:2016dpx}. 

In cross-correlation method, one can define the product of two outputs as the correlator, then the measurement can be obtained by integrating the correlator over time:
\bea
\label{eq:S_IJ}
\nn
S_{IJ}(t_{0})&=&\int_{t_{0}-T/2}^{t_{0}+T/2}{\rm d}t\int_{t_{0}-T/2}^{t_{0}+T/2}{\rm d}t'
s_{I}(t)s_{J}(t')Q_{IJ}(t-t',t_{0})\\
\nn
&\approx&
\int_{t_{0}-T/2}^{t_{0}+T/2}{\rm d}t
\int_{-\infty}^{\infty}{\rm d}f\int_{-\infty}^{\infty}{\rm d}f'\,
\widetilde{s}_{I}(f,t_{0})\widetilde{s}_{J}^{*}(f',t_{0})\\
&\quad&\times
\frac{P^{*}_{{\rm h}_{IJ}}(f,t_{0})}{P_{{\rm n}_{I}}(f)P_{{\rm n}_{J}}(f)W_{IJ}(f,t_{0})}e^{-{\rm i}2\pi(f-f')t},
\eea
where $P_{{\rm n}}$ is the noise \ac{PSD}, and the correction function 
\bea
\label{eq:W_IJ}
\nn
W_{IJ}(f,t_{0})
&=&1+\frac{P_{{\rm h}_{I}}(f,t_{0})P_{{\rm n}_{J}}(f)+P_{{\rm h}_{J}}(f,t_{0})P_{{\rm n}_{I}}(f)}{P_{{\rm n}_{I}}(f)P_{{\rm n}_{J}}(f)}\\
&&+
\frac{P_{{\rm h}_{I}}(f,t_{0})P_{{\rm h}_{J}}(f,t_{0})+
|P_{{\rm h}_{IJ}}(f,t_{0})|^{2}}{P_{{\rm n}_{I}}(f)P_{{\rm n}_{J}}(f)}.
\eea
Under the assumption that \ac{SGWB} signal and channel noise are stationary in the time interval $[t_{0}-T/2,t_{0}+T/2]$, the filter function $Q_{IJ}(t,t')=Q_{IJ}(t-t')$. 

In terms of the expectation value and variance of the measurement
\bea
\nn
\mu(t_{0})&=&\langle S_{IJ}(t_{0})\rangle,\\
\sigma^{2}(t_{0})&=&
\langle S_{IJ}(t_{0})S_{IJ}(\eta_{0})\rangle-\langle S_{IJ}(t_{0})\rangle
\langle S_{IJ}(\eta_{0})\rangle,
\eea
the \ac{SNR} can be obtained by
\be
\label{eq:snr_cc}
\rho(t_{0})=\frac{\mu(t_{0})}{\sigma(t_{0})}=
\sqrt{2\,T\int_{f_{\rm min}}^{f_{\rm max}}{\rm d}f\,
\frac{|P_{{\rm h}_{IJ}}(f,t_{0})|^{2}}{P_{{\rm n}_{I}}(f)P_{{\rm n}_{J}}(f)W_{IJ}(f,t_{0})}}.
\ee

For a single TianQin-like detector, the correlation between the \ac{SGWB} signal of noise-independent channels cancels~\cite{Cutler:1997ta,Adams:2010vc}. 
Thus the cross-correlation method will fall, and the null-channel method need to be introduced. 

For the channel group $\rm AET$, one can construct the correlator for null-channel method by
\be
s_{0}(t,t')=\sum_{I=\rm A,E}\bigg[s_{I}(t)s_{I}(t')-\frac{1}{2}\int_{-\infty}^{\infty}{\rm d}f
\,e^{{\rm i}2\pi f(t-t')}P_{{\rm n}_{I}}(f)\bigg],
\ee
where the noise \ac{PSD} of $\rm A/E$ channels can be monitored by the null channel $\rm T$. 
Similar to \eq{eq:S_IJ}, we can obtain the measurement by
\bw
\bea
\nn
K(t_{0})&=&\int_{t_{0}-T/2}^{t_{0}+T/2}{\rm d}t\int_{t_{0}-T/2}^{t_{0}+T/2}{\rm d}t'
s_{0}(t,t')Q_{II}(t-t',t_{0})\\
&\approx&\sum_{I=\rm A,E}
\int_{t_{0}-T/2}^{t_{0}+T/2}{\rm d}t
\int_{-\infty}^{\infty}{\rm d}f\int_{-\infty}^{\infty}{\rm d}f'
\,\bigg[\widetilde{s}_{I}(f,t_{0})
\widetilde{s}_{I}^{*}(f',t_{0})-\frac{1}{2}P_{{\rm n}_{I}}(f)\bigg]
\frac{P_{{\rm h}_{I}}(f,t_{0})e^{-{\rm i}2\pi(f-f')t_{0}}}{P^{2}_{{\rm n}_{I}}(f)W_{I}(f,t_{0})},
\eea
\ew
where the correction function
\bea
W_{I}(f,t_{0})
&=&\bigg(1+\frac{P_{{\rm h}_{I}}(f,t_{0})}{P_{{\rm n}_{I}}(f)}\bigg)^{2}.
\eea
For a symmetric scenario, one can further assume
\bea
\nn
P_{{\rm h}_{\rm A}}(f,t_{0})&=&P_{{\rm h}_{\rm E}}(f,t_{0}),\\
P_{{\rm n}_{\rm A}}(f)&=&P_{{\rm n}_{\rm E}}(f).
\eea
Then the \ac{SNR} of null-channel method is given by
\bea
\label{eq:snr_null}
\nn
\rho(t_{0})&=&\frac{\langle K(t_{0})\rangle}{\sqrt{\langle K(t_{0})K(\eta_{0})\rangle-\langle K(t_{0})\rangle\langle K(\eta_{0})\rangle}}\\
&=&
\sqrt{2\,T\int_{f_{\rm min}}^{f_{\rm max}}{\rm d}f\,
\frac{P^{2}_{{\rm h}_{I}}(f,t_{0})}{P^{2}_{{\rm n}_{I}}(f)W_{I}(f,t_{0})}}.
\eea
By setting $I=J$, the \ac{SNR} of cross-correlation method (i.e., \eq{eq:snr_cc}) returns to that of null-channel method (i.e., \eq{eq:snr_null}). 
More details are shown in Appendix~\ref{appen:Q_W}. 

As the \ac{SNR} is proportional to $\sqrt{T}$, we can accumulate a sufficiently high \ac{SNR} by correlating plenty of data sets. 
For the total correlation time $T_{\rm tot}=nT$, the \ac{SNR}
\bea
\label{eq:SNR_tot}
\nn
\rho&=&
\sqrt{\sum_{t_{0}=0}^{(n-1)T}\rho^{2}(t_{0})}\\
\nn
&=&
\sqrt{2\,T\int_{f_{\rm min}}^{f_{\rm max}}{\rm d}f\,\sum_{t_{0}=0}^{(n-1)T}
\frac{|P_{{\rm h}_{IJ}}(f,t_{0})|^{2}}{P_{{\rm n}_{I}}(f)P_{{\rm n}_{J}}(f)W_{IJ}(f,t_{0})}}.\\
\eea
When \ac{SGWB} signal is much stronger than channel noise, the total \ac{SNR} 
\bea
\label{eq:rho_large}
\nn
\rho&=&\sqrt{2\,T\int_{f_{\rm min}}^{f_{\rm max}}{\rm d}f\,\sum_{t_{0}=0}^{(n-1)T}
\frac{1}{\frac{P_{{\rm h}_{I}}(f,t_{0})P_{{\rm h}_{J}}(f,t_{0})}{|P_{{\rm h}_{IJ}}(f,t_{0})|^{2}}+(1-\delta_{IJ})}}\\
&\leq&\sqrt{(1+\delta_{IJ})T_{\rm tot}(f_{\rm max}-f_{\rm min})},
\eea
where $|P_{{\rm h}_{IJ}}(f,t_{0})|^{2}\leq P_{{\rm h}_{I}}(f,t_{0})P_{{\rm h}_{J}}(f,t_{0})$. 
We can find that the \ac{SNR} is limited to a certain value,  which is determined by three factors: (i) the type of detection method; (ii) the correlation time $T_{\rm tot}$; (iii) detection frequency band $[f_{\rm min},f_{\rm max}]$. 
On the contrary, when \ac{SGWB} is much weaker than channel noise, the correction function $W_{IJ}(f,t_{0})\to 1$. 
In this case, the \ac{SNR} for an isotropic \ac{SGWB} can be simplified to the following form: 
\be
\label{eq:snr_iso}
\rho
=\sqrt{2\,T_{\rm tot}\int_{f_{\rm min}}^{f_{\rm max}}{\rm d}f\,
\frac{\big[\bar{\Gamma}_{IJ}(f)
S_{\rm h}(f)\big]^{2}}{P_{{\rm n}_{I}}(f)P_{{\rm n}_{J}}(f)}}.
\ee
When \ac{SNR} exceeds the preset threshold, the \ac{SGWB} detection will be announced. 

As for the \ac{SNR} threshold, the preliminary result is provided by Ref.~\cite{Allen:1997ad}:
\be
\label{eq:rho_th}
{\rm {\rho}_{\rm thr}}=\sqrt{2}[{\rm erfc}^{-1}(2\alpha)-{\rm erfc}^{-1}(2\gamma)],
\ee
where the false alarm rate $\alpha$ and the detection rate $\gamma$ are set in advance. 
For example, when choosing $\alpha=0.1$ and $\gamma=0.9$, the \ac{SNR} threshold ${\rho}_{\rm thr}=2.56$; or ${\rho}_{\rm thr}=3.30$ with $\alpha=0.05$ and $\gamma=0.95$. 
Because the lower false alarm rate and the higher detection rate imply that the detector is more sensitive to \acp{GW}, the higher \ac{SNR} threshold is required.

%==========The sensitivity curve for SGWB detection===========%
\section{Sensitivity curve}\label{sec:sensitivity}
We can further demonstrate the detection capability of the detector through the sensitivity curve. 
The {\it sky-averaged} sensitivity curve can be directly obtained based on the response to \ac{SGWB} and noise lever of the channel~\cite{Cornish:2001bb}:
\be
\label{eq:S_nI}
S_{{\rm n}_{I}}(f)=\frac{P_{{\rm n}_{I}}(f)}{\mathcal{R}_{I}(f)},
\ee
which is applied to an isotropic \ac{SGWB}.

As for two channels, the effective sensitivity curve is defined as~\cite{Thrane:2013oya}
\be
\label{eq:S_nIJ}
S_{{\rm n}_{IJ}}(f)
=\frac{\sqrt{P_{{\rm n}_{I}}(f)P_{{\rm n}_{J}}(f)}}{\bar{\Gamma}_{IJ}(f)},
\ee
which reduces to \eq{eq:S_nI} when $I=J$. Then, \eq{eq:snr_iso} can be further simplified:
\be
\label{eq:snr_re}
\rho
=\sqrt{2\,T_{\rm tot}\int_{f_{\rm min}}^{f_{\rm max}}{\rm d}f\,
\bigg[\frac{S_{\rm h}(f)}{S_{{\rm n}_{IJ}}(f)}\bigg]^{2}}.
\ee
Meanwhile, the corresponding energy spectral density $\Omega_{{\rm n}_{IJ}}$ can be convert to $S_{{\rm n}_{IJ}}$ through
\be
\label{eq:Omega_n}
\Omega_{{\rm n}_{IJ}}(f)=
\frac{2\pi^{2}}{3H_{0}^{2}}f^{3}S_{{\rm n}_{IJ}}(f).
\ee	

As mentioned above, the correlation time and frequency band of the detector have a significant impact on \ac{SGWB} detection. Therefore, Thrane et.al~\cite{Thrane:2013oya} proposed the \ac{PLIS} curve, which is applied to the power-law \ac{SGWB} with the following form:
\be
\label{eq:Omegaform}
\Omega_{\rm gw}(f)=\Omega_{0}(\epsilon)(f/f_{\rm ref})^{\epsilon}|_{\epsilon=\epsilon_{0}},
\ee
where $\Omega_{0}$ is related to the index $\epsilon$, and the reference frequency $f_{\rm ref}$ is arbitrary. 

By combining \eq{eq:snr_re}-\eq{eq:Omegaform},
\be
\Omega_{0}(\epsilon)=\rho_{\rm thr}\left[2\,T_{\rm tot}
\int_{f_{\rm min}}^{f_{\rm max}}{\rm d}f
\frac{(f/f_{\rm ref})^{2\epsilon}}{\Omega_{{\rm n}_{IJ}}^{2}(f)}\right]^{-1/2},
\ee
where the \ac{SNR} is set to threshold $\rho_{\rm thr}$. 
For each frequency, we can obtain the maximum $\Omega_{\rm gw}$ with a specific index $\epsilon$ to generate the \ac{PLIS} curve:
\be
\label{eq:Omega_PLI}
\Omega_{\rm PLIS}(f)={\rm max}_{\epsilon}[\Omega_{0}(\epsilon)(f/f_{\rm ref})^{\epsilon}].
\ee
The \ac{PLIS} curve (on the log-log plot) specifies the envelope of power-law \acp{SGWB}, of which the \ac{SNR} is equal to the preset \ac{SNR} threshold. 
Once a power-law \ac{SGWB} spectrum is somewhere above the \ac{PLIS} curve, the \ac{SGWB} is expected to be detected, and vice versa. 
It is explicit to determine whether the \ac{SGWB} can be detected by the detector configuration. 

\begin{figure}[t]
\centering
\includegraphics[height=6cm]{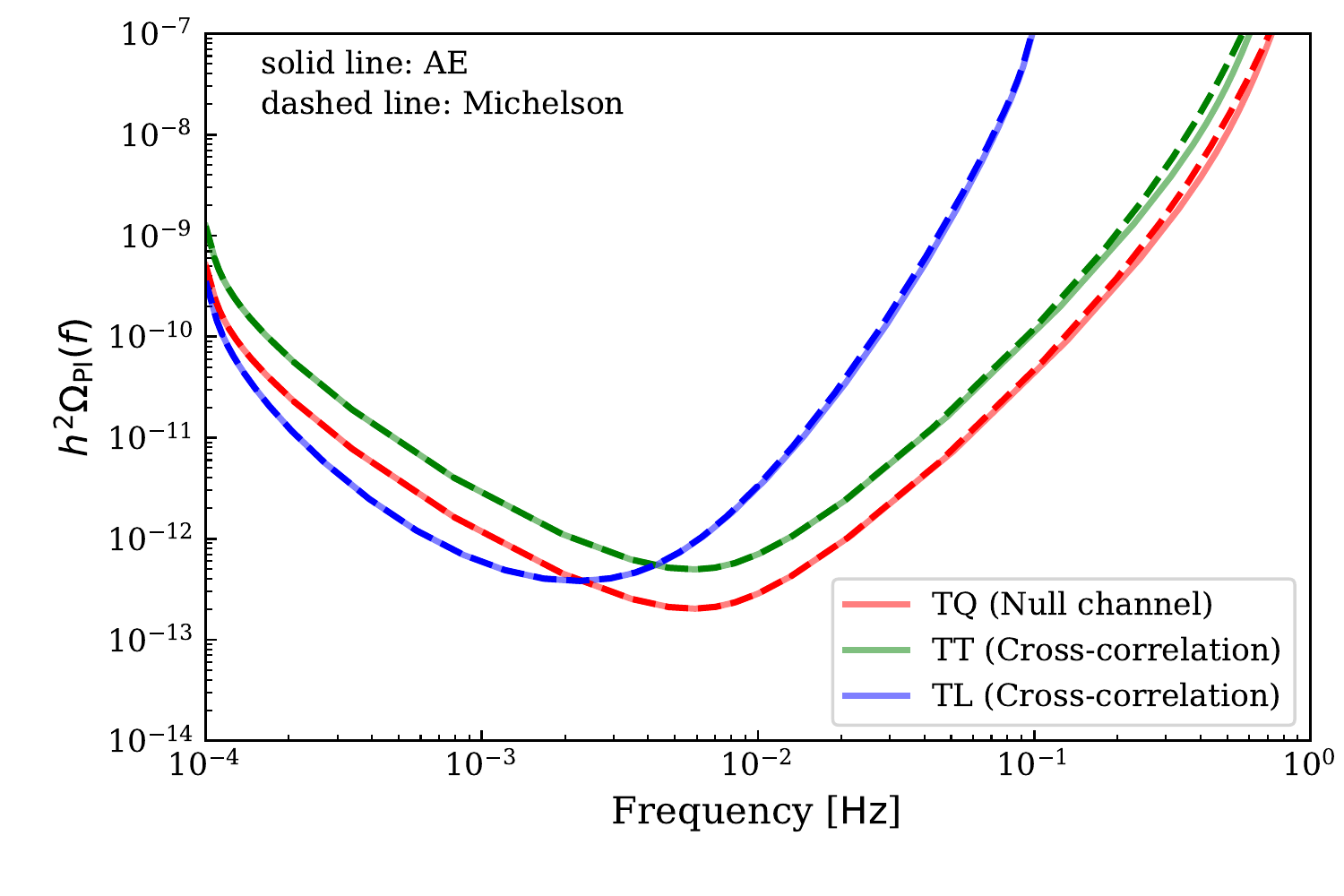}
\caption{PL sensitivity curves for different detector configurations. We set the \ac{SNR} threshold to 3.30 and consider that during the one-year operating time, the correlation time of TianQin and TianQin + LISA is half a year, compared to four months for TianQin I+II.}
\label{fig:PI_curve}
\end{figure}

We show the \ac{PLIS} curves of TianQin, TianQin I+II, TianQin + LISA in \fig{fig:PI_curve}. 
The solid and dashed lines are the results for the equal-arm Michelson and $\rm A/E$ channels, and both of them basically coincide. 
The result implies that when $\rm A/E$ channels are employed instead of the equal-arm Michelson, the detection capability of the detector configuration is nearly unchanged. 
%Since $\rm A/E$ channels contain multiple laser interference sites, there still exists a slight gap at high frequencies. 

The spectrum of astrophysical \ac{SGWB} is usually in power-law form, which is not true for the cosmological \ac{SGWB}, such as the first-order \ac{PT}. 
Therefore, Schmitz et.al~\cite{Schmitz:2020syl} proposed the \ac{PIS} curve. 

The energy spectral density of the first-order \ac{PT} is expressed as
\be
\Omega_{\rm gw}(f)=\Omega_{\rm gw}^{\rm peak}(\{p_{i}\})S(f,f_{\rm peak}),
\ee
where $\Omega_{\rm gw}^{\rm peak}(\{p_{i}\})$ is the peak amplitude at the peak frequency $f_{\rm peak}$, the spectral function $S(f,f_{\rm peak})$ depends on the cosmological model. 
Then through the definition of \ac{PIS} curve
\be
\Omega_{\rm PIS}(f_{\rm peak})=\left[2\,T_{\rm tot}
\int_{f_{\rm min}}^{f_{\rm max}}{\rm d}f
\bigg(\frac{S(f,f_{\rm peak})}{\Omega_{{\rm n}_{IJ}}(f)}\bigg)^{2}\right]^{-1/2},
\ee 
the \ac{SNR} can be obtained by
\be
\rho=\frac{\Omega_{\rm gw}^{\rm peak}(\{p_{i}\})}{\Omega_{\rm PIS}(f_{\rm peak})}. 
\ee
Once selecting a specific parameter group $\{p_{i}\}$, the peak frequency $f_{\rm peak}$ and peak amplitude $\Omega_{\rm gw}^{\rm peak}$ are fixed. 
If the peak amplitude is above $\rho_{\rm thr}$ times of the \ac{PIS} curve, the detection of cosmological \ac{PT} will be claimed. 
\begin{figure}[t]
\centering
\includegraphics[height=6cm]{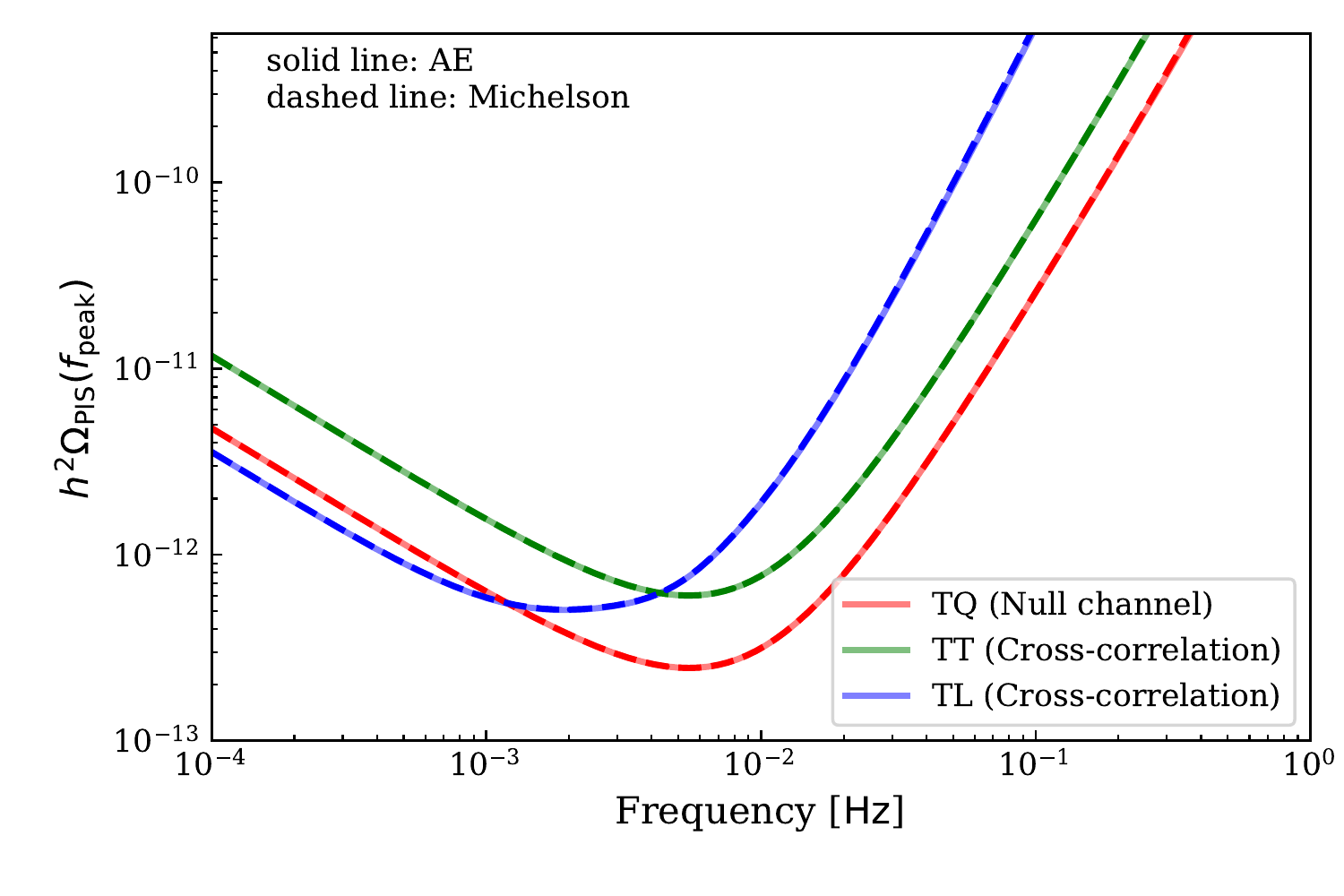}
\caption{\ac{PIS} curves of TianQin (red), TianQin I+II (green) and TianQin +LISA (blue) for $\rm A/E$ channels (solid line) and equal-arm Michelson (dashed line). The setting of the correlation time for the configuration is the same as when plotting the \ac{PLIS} curve.}
\label{fig:PIS_curve}
\end{figure}

In \fig{fig:PIS_curve}, we show the \ac{PIS} curves for a certain spectral function:
\be
S(f,f_{\rm peak})
=\frac{3.8(f/f_{\rm peak})^{2.9}}{1+2.9(f/f_{\rm peak})^{3.8}}.
\ee
Similar to the \ac{PLIS} curve, the \ac{PIS} curves of equal-arm and $\rm A/E$ channels are mostly the same. 

So far, this paper has discussed three types of sensitivity curves. 
Sky-averaged sensitivity curve is independent of correlation time and detection frequency band, and by considering the impact of the two factors on \ac{SGWB} detection, \ac{PLIS} curve and \ac{PIS} curve are proposed. 
It is straightforward to determine whether \ac{SGWB} can be detected by the last two sensitivity curves. 
Besides, the \ac{PLIS} curve works for the pow-law \ac{SGWB}, and the \ac{PLIS} curve is applied to the \ac{SGWB} with pre-known spectrum. 

\section{Conclusion}\label{sec:conclusion}
%\ac{SGWB} detection contributes to a deeper understanding of nearby compact objects and early Universe. 
In this work, we have analyzed the impacts of $\rm A/E$ channels on \ac{SGWB} detection for space-borne detectors. 
We first selected the reference interference site for $\rm A/E$ channels consisting multiple interference sites. 
In this way, it is clear to define the detector separation through the distance between reference interference sites, which works well for any channel. 
By means of the reference interference site, we derived the \ac{ORF} that is valid across all frequency bands. 
For TianQin, TianQin I+II and TianQin + LISA, we calculated the \acp{ORF} for the equal-arm Michelson and $\rm A/E$ channels, respectively. 
In addition to \ac{ORF}, the noise level, correlation time and frequency band of the detector also need to be folded into detection sensitivity, according to which we plotted the corresponding sensitivity curves to assess the detection capabilities. 

In a certain detector configuration, the \ac{ORF} and noise \ac{PSD} of A/E channels differ from those of the equal-arm Michelson, but their differences share the same coefficient $6\sin^2(f/f_{\ast})$ under the low-frequency approximation. 
Since the detection sensitivity is determined by the ratio of \ac{ORF} and noise \ac{PSD}, the detection sensitivity of the $\rm A/E$ channels and the equal-arm Michelson are basically the same. 
We have shown that the difference between A/E channels and the equal-arm Michelson stems from the extra phase related to the position difference.
For the low-frequency approximation to be valid, the wavelength of the \ac{GW} should be longer than the armlength, which makes the extra phase negligible.

However, the configuration design can affect the correlation between detectors. 
In order to obtain optimal \ac{ORF}, the configuration design is well worth a discussion. 
Meanwhile, the constructions of equal-arm Michelson and $\rm A/E$ channels are different. 
Therefore, the configuration design corresponding to the optimal \ac{ORF} for $\rm A/E$ channels may differ from that for the equal-arm Michelson. 
TianQin I+II is one such configuration, where the launch time difference between TianQin and TianQin II is the key factor of configuration design. 
If one expects the optimal \ac{ORF} of TianQin I+II in \ac{SGWB} detection, then the launch time difference should be set to an integer and a semi-integer multiple of the orbital period for the equal-arm Michelson and $\rm A/E$ channels, respectively. 

Recently, Bartolo et al.~\cite{LISACosmologyWorkingGroup:2022kbp}~reviewed the sensitivity curve for the detection of the anisotropic \ac{SGWB} by a single LISA. 
Unlike for the isotropic \ac{SGWB}, where only the $\rm A/E$ channels and the equal-arm Michelson contribute to the detection sensitivity, for the anisotropic \ac{SGWB} the null channel $\rm T$ also has a non-negligible impact on the sensitivity. 
Therefore, if only considering the A/E channels and the equal-arm Michelson, the detection sensitivities for the anisotropic \ac{SGWB} are different. 
However, by incorporating the null channel, the detection sensitivities for $\rm A/E$ and $\rm T$, as well as the equal-arm Michelson and $\rm T$ for the anisotropic \ac{SGWB} will be equivalent.

\begin{acknowledgments}
This work has been supported by the Guangdong Major Project of Basic and Applied Basic Research (Grant No. 2019B030302001), the National Key Research and Development Program of China (No. 2020YFC2201400), the Natural Science Foundation of China (Grants No. 12173104), and the Natural Science
Foundation of Guangdong Province of China (Grant No.
2022A1515011862). 
We also thank Shun-Jia Huang, Xiang-Yu Lyu, Jianwei Mei for helpful discussions.
\end{acknowledgments}

\appendix
\bw
\section{Response to Gravitational-wave for equal-arm Michelson}\label{appen:channel_response}
\begin{figure}[t]
\centering
\includegraphics[height=5cm]{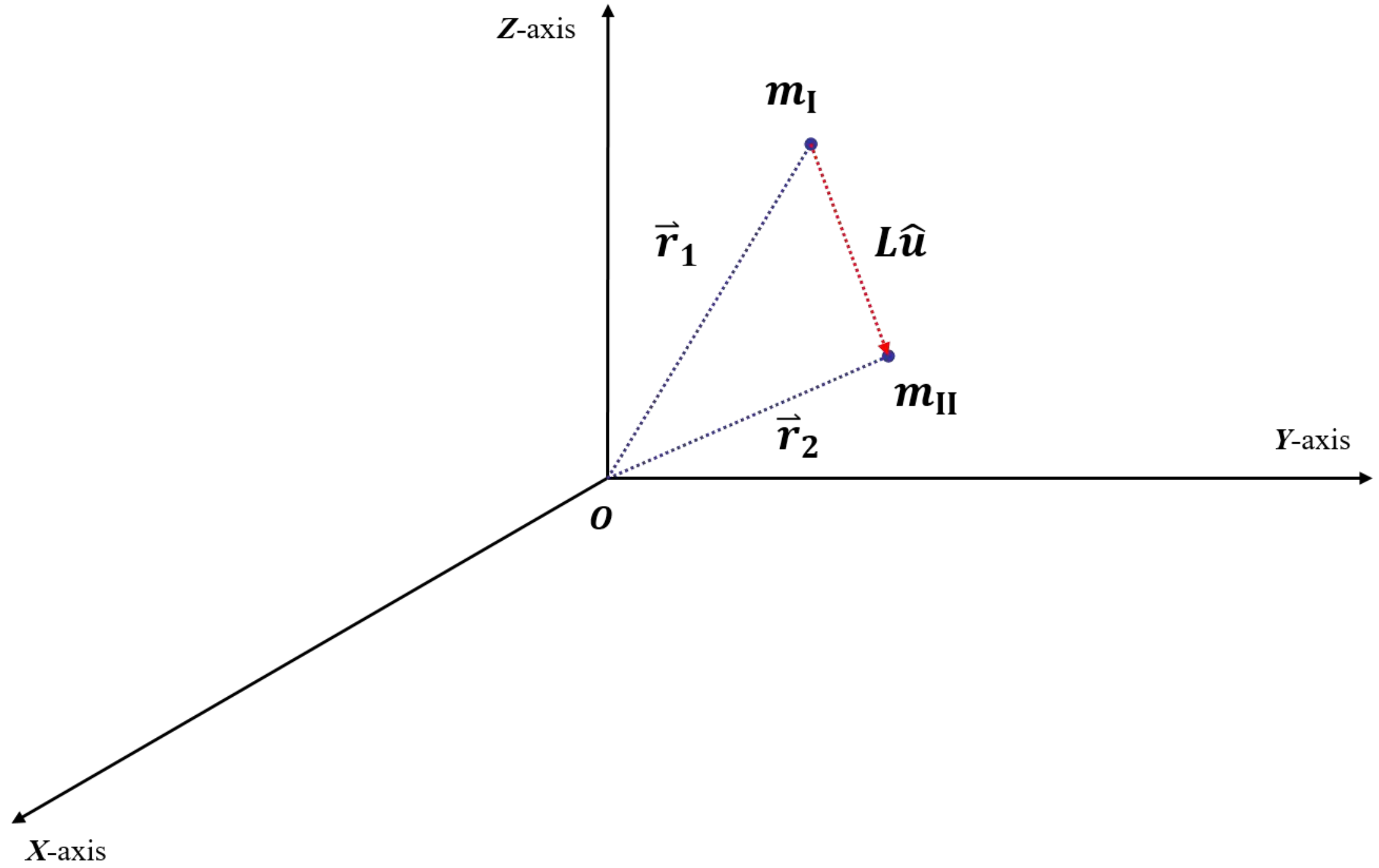}
\caption{Schematic diagram of the one-way tracking.}
\label{fig:one-way}
\end{figure}

In \fig{fig:one-way}, $\vec{r}_{1}$ and $\vec{r}_{2}$ label the position vectors of test mass $m_{\rm I}$ and $m_{\rm II}$, respectively. 
$L$ is armlength and $\hat{u}$ denotes the unit vector of the one-way tracking. 
At the moment $t_{0}$, the \ac{GW} signal of one-way tracking can be expressed as~\cite{Abramovici:1992ah,Romano:2016dpx}
\be
h_{\rm I}(t,t_{0})=
\frac{\delta L(t_{0})}{L}=
\frac{1}{L}\int_{0}^{L}{\rm d}s\,\frac{u^{a}(t_{0})u^{b}(t_{0})}{2}h_{ab}[t(s),\vec{x}(s)],
\ee
where $s$ is the actual path of photon, and under the 0th-order approximation:
\bea
\nn
t(s)&=&(t_{0}-L/c)+s/c,\\
\vec{x}(s)&=&\vec{r}_{1}+s\hat{u}.
\eea
In terms of \eq{eq:h_ab}, the one-way tracking signal of \ac{SGWB} is a collection of \ac{GW} signals from all directions:
\bea
\label{eq:ht_1w}
\nn
h_{\rm I}(t,t_{0})&=&
\frac{1}{L}\int_{-\infty}^{\infty}{\rm d}f\,\int_{S^{2}}{\rm d}\hat{\Omega}_{\hat{k}}
\frac{u^{a}(t_{0})u^{b}(t_{0})}{2}h_{ab}(f,\hat{k})
e^{{\rm i}2\pi f[t-\frac{L+\hat{k}\cdot(\vec{r}_{2}-L\hat{u})}{c}]}
\frac{c\cdot e^{-\frac{{\rm i}2\pi fs}{c}(\hat{k}\cdot\hat{u}(t_{0})-1)}}{-{\rm i}2\pi f(\hat{k}\cdot\hat{u}(t_{0})-1)}
\bigg\vert_{s=0}^{s=L}\\
\nn
&=&\int_{-\infty}^{\infty}{\rm d}f\,
\int_{S^{2}}{\rm d}\hat{\Omega}_{\hat{k}}
F_{\rm I}^{ab}(f,\hat{k},t_{0})h_{ab}(f,\hat{k})
e^{{\rm i}2\pi f(t-\frac{\hat{k}\cdot\vec{r}_{2}}{c})},
\eea
then the frequency domain signal
\be
\label{eq:hf_1w}
h_{\rm I}(f,t_{0})=\int_{S^{2}}{\rm d}\hat{\Omega}_{\hat{k}}
F_{\rm I}^{ab}(f,\hat{k},t_{0})h_{ab}(f,\hat{k})
e^{-{\rm i}2\pi f\hat{k}\cdot\vec{r}_{2}/c}.
\ee
In \eq{eq:hf_1w}, the one-way tracking response function
\be
\label{eq:F_1w}
F_{\rm I}^{ab}(f,\hat{k},t_{0})
=\frac{u^{a}(t_{0})u^{b}(t_{0})}{2}\mathcal{T}_{\rm I}(f,\hat{k},t_{0})
\ee
and the strain transfer function
\be
\mathcal{T}_{\rm I}(f,\hat{k},t_{0})
={\rm sinc}\big[\frac{f}{2f_{\ast}}[1-\hat{k}\cdot\hat{u}(t_{0})]\big]
e^{-i\frac{f}{2f_{\ast}}[1-\hat{k}\cdot\hat{u}(t_{0})]}
\ee
with the characteristic frequency $f_{\ast}=c/(2\pi L)$. 

For the two-way tracking where the photon returns to $m_{\rm I}$ after a one-way tracking, the \ac{SGWB} signal
\bea
\label{eq:ht_2w}
\nn
h_{\rm II}(t,t_{0})&=&\frac{1}{2L}
\int_{0}^{L}{\rm d}s\,\frac{u^{a}(t_{0})u^{b}(t_{0})}{2}
\bigg[h_{ab}\bigg(t-\frac{2L+\hat{k}\cdot\vec{r}_{1}+[\hat{k}\cdot\hat{u}(t_{0})-1]s}{c}\bigg)+
h_{ab}\bigg(t-\frac{L+\hat{k}\cdot\vec{r}_{2}+[\hat{k}\cdot-\hat{u}(t_{0})-1]s}{c}\bigg)\bigg]\\
&=&\int_{-\infty}^{\infty}{\rm d}f\,
F_{\rm II}^{ab}(f,\hat{k},t_{0})h_{ab}(f,\hat{k})
e^{{\rm i}2\pi f(t-\frac{\hat{k}\cdot\vec{r}_{1}}{c})}.
\eea
Note that in the above formula, the length of links is assumed to not change over time. 

Then the corresponding frequency domain signal
\be
\label{eq:hf_2w}
h_{\rm II}(f,t_{0})
=\int_{-\infty}^{\infty}{\rm d}f\,
F_{\rm II}^{ab}(f,\hat{k},t_{0})h_{ab}(f,\hat{k})
e^{-{\rm i}2\pi f\hat{k}\cdot\vec{r}_{1}/c},
\ee
where the response function
\be
\label{eq:F_2w}
F_{\rm II}^{ab}(f,\hat{k},t_{0})=
\frac{1}{2}\mathcal{T}_{\rm II}(f,\hat{k},t_{0})u^{a}(t_{0})u^{b}(t_{0})
\ee
with the strain transfer function
\bea
\mathcal{T}_{\rm II}(f,\hat{k},t_{0})
&=&\frac{1}{2}
\bigg[{\rm sinc}
\big[\frac{f}{2f_{\ast}}\big(1-\hat{k}\cdot\hat{u}(t_{0})\big)\big]
e^{-i\frac{f}{2f_{\ast}}[3+\hat{k}\cdot\hat{u}(t_{0})]}
+{\rm sinc}
\big[\frac{f}{2f_{\ast}}\big(1+\hat{k}\cdot\hat{u}(t_{0})\big)\big]
e^{-i\frac{f}{2f_{\ast}}[1+\hat{k}\cdot\hat{u}(t_{0})]}\bigg].
\eea
For the record, the phase terms in \eq{eq:hf_1w} and \eq{eq:hf_2w} are related to the \ac{GW} measurement location. 
%Since \eq{eq:F_1w} and \eq{eq:F_2w} characterize the \ac{GW} response for the same link, they reduce to $u^{a}(t_{0})u^{b}(t_{0})/2$ under the low-frequency approximation (i.e., $f\ll f_{\ast}$). 

Since the equal-arm Michelson consists of two different two-way trackings, the response function can be written as
\be
\label{eq:Fab}
F_{\rm M}^{ab}(f,\hat{k},t_{0})
=F_{\rm II}^{ab}(f,\hat{k},\hat{u}(t_{0}))-F_{\rm II}^{ab}[f,\hat{k},\hat{v}(t_{0})],
\ee
where $\hat{u}$ and $\hat{v}$ are the unit vectors of the arms. Under the low-frequency approximation, the response function
\be
F_{\rm M}^{ab}(f,\hat{k},t_{0})
=\frac{1}{2}[u^{a}(t_{0})u^{b}(t_{0})-v^{a}(t_{0})v^{b}(t_{0})].
\ee 

\section{ORF of channel group}\label{appen:ORF_Michelson}
\begin{figure}[h]
\centering
\includegraphics[height=6cm]{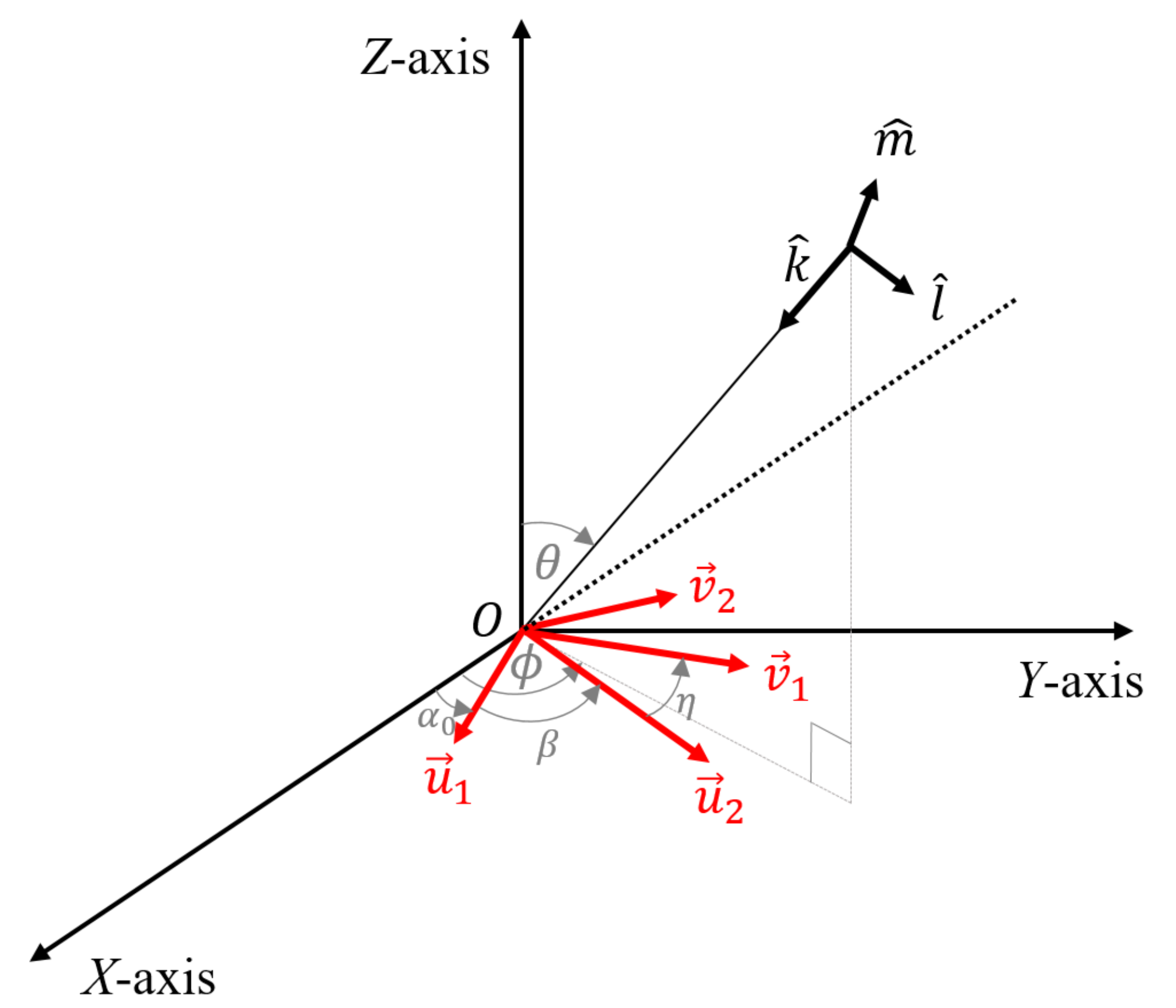}
\caption{{Detector coordinate}}
\label{fig:D_c}
\end{figure}
As shown in \fig{fig:D_c}, we can construct a set of orthogonal bases:
\begin{eqnarray}
\hat{k}&=&(-\sin\theta \cos\phi,-\sin\theta \sin\phi,-\cos\theta),\\
\hat{l}&=&(\cos\theta \cos\phi,\cos\theta \sin\phi,-\sin\theta),\\
\hat{m}&=&(\sin\phi,-\cos\phi,0),
\end{eqnarray}
where $\hat{k}$ is the propagation direction of the \ac{GW}, and $\hat{m}$ is perpendicular to $Z$-axis. 

In terms of $\hat{l}$ and $\hat{m}$, we can further construct the polarization tensor:
\begin{eqnarray}
e^{+}_{ij}(\hat{n})&=&\hat{l}_{i}\hat{l}_{j}-\hat{m}_{i}\hat{m}_{j},\\
e^{\times}_{ij}(\hat{n})&=&\hat{l}_{i}\hat{m}_{j}+\hat{l}_{j}\hat{m}_{i}.
\end{eqnarray} 
Then the analytical expression of the polarization tensor can be obtained:
\begin{eqnarray}
\nn
e^{+}(\hat{k})&=&{
\left[\begin{array}{ccc}
	\cos^{2}\theta \cos^{2}\phi-\sin^{2}\phi &
	\frac{1}{2}\sin2\phi(1+\cos^{2}\theta) &
	-\frac{1}{2}\sin2\theta \cos\phi\\
	\frac{1}{2}\sin2\phi(1+cos^{2}\theta) & \sin^{2}\phi \cos^{2}\theta-\cos^{2}\phi & -\frac{1}{2}\sin2\theta \sin\phi\\
	-\frac{1}{2}\sin2\theta \cos\phi & -\frac{1}{2}\sin2\theta
	\sin\phi & \sin^{2}\theta
\end{array}\right],}\\
e^{\times}(\hat{k})&=&{
	\left[\begin{array}{ccc}
	\sin2\phi\cos\theta & -\cos2\phi \cos\theta & -\sin\phi \sin\theta\\
	-\cos2\phi \cos\theta & -\sin2\phi \cos\theta & \cos\phi \sin\theta\\
	-\sin\phi \sin\theta & \cos\phi \sin\theta & 0
\end{array}\right].}
\end{eqnarray}

On the $X$-$Y$ plane, we place two equal-arm Michelson channels with opening angle $\beta$ between arms. 
The unit arm vectors of the first one $\rm M_{1}$
\bea
\nn
\hat{u}_{1}&=&[\cos\alpha_{0},\sin\alpha_{0},0\big],\\
\hat{u}_{2}&=&[\cos(\alpha_{0}+\beta),\sin(\alpha_{0}+\beta),0],
\eea
and of the second one $\rm M_{2}$
\bea
\nn
\vec{v}_{1}&=&[\cos(\alpha_{0}+\beta+\eta),\sin(\alpha_{0}+\beta+\eta),0]\\
\vec{v}_{2}&=&[\cos(\alpha_{0}+2\beta+\eta),\sin(\alpha_{0}+2\beta+\eta),0].
\eea

Without loss off generality, one can set $\alpha_{0}=0$ to simplify the calculation, and under the low-frequency approximate,
\bea
\nn
F_{\rm M_{1}}
&=\frac{1}{2}({u}_{1}\otimes{u}_{1}-{u}_{2}\otimes{u}_{2})=&
{\frac{\sin\beta}{2}\left[\begin{array}{ccc}
		\sin\beta          & -\cos\beta      & 0\\
		-\cos\beta         & -\sin\beta      & 0\\
		0                  & 0               & 0
	\end{array}\right],}\\
F_{\rm M_{2}}
&=\frac{1}{2}({v}_{1}\otimes{v}_{1}-{v}_{2}\otimes{v}_{2})=&
{\frac{\sin\beta}{2}\left[\begin{array}{ccc}
		\sin(3\beta+2\eta)         & -\cos(3\beta+2\eta)  & 0\\
		-\cos(3\beta+2\eta)        & -\sin(3\beta+2\eta)  & 0\\
		0                          & 0                    & 0
	\end{array}\right].}
\eea
Then through
\bea
\nn
F_{\rm M_{1}}^{+}(0,\hat{k})&=&F_{\rm M_{1}}^{ab}e_{ab}^{+}(\hat{k})
=\frac{3+\cos2\theta}{4}\sin\beta \sin(\beta-2 \phi), \\
\nn
F_{\rm M_{1}}^{\times}(0,\hat{k})&=&F_{\rm M_{1}}^{ab}e_{ab}^{\times}(\hat{k})
=\sin\beta \cos\theta \cos(\beta-2 \phi), \\
\nn
F_{\rm M_{2}}^{+}(0,\hat{k})&=&F_{\rm M_{1}}^{ab}e_{ab}^{+}(\hat{k})
=\frac{3+\cos2\theta}{4}\sin\beta\sin(3 \beta+2 \eta-2 \phi),  \\
F_{\rm M_{2}}^{\times}(0,\hat{k})&=&F_{\rm M_{1}}^{ab}e_{ab}^{\times}(\hat{k})
=\sin\beta \cos\theta \cos(3 \beta+2 \eta-2 \phi),
\eea
we can calculate the transfer function and \ac{ORF}:
\bea
\nn
\mathcal{R}_{{\rm M}_{1}}(0)
&=&\frac{1}{8\pi}\sum_{P=+,\times}\int_{S^{2}}{\rm d}\hat{\Omega}_{\hat{k}}
F_{\rm M_{1}}^{P}(0,\hat{k})F_{\rm M_{1}}^{P*}(0,\hat{k})
=\frac{\sin^{2}\beta}{5}=\mathcal{R}_{{\rm M}_{2}}(0),\\
\Gamma_{{\rm M}_{12}}(0)
&=&\frac{1}{8\pi}\sum_{P=+,\times}\int_{S^{2}}{\rm d}\hat{\Omega}_{\hat{k}}
F_{\rm M_{1}}^{P}(0,\hat{k})F_{\rm M_{2}}^{P*}(0,\hat{k})
=\frac{\sin^{2}\beta \cos[2(\beta+\eta)]}{5}.
\eea

For the equal-arm Michelson channels built in TianQin, 
$\beta=\pi/3$ and $\beta+\eta=2\pi/3$, then
\be
\mathcal{R}_{{\rm M}_{1}}(0)=-2\Gamma_{{\rm M}_{12}}(0).
\ee
In terms of \eq{eq:F_X} and \eq{eq:F_YZ}, 
\be
\lim_{f\to 0} \frac{\mathcal{R}_{\rm X}(f)}{\Gamma_{\rm XY}(f)}
=\lim_{f\to 0} \frac{\mathcal{R}_{{\rm M}_{1}}(f)}{\Gamma_{{\rm M}_{12}}(f)}=-2.
\ee
Besides, for an regular triangle detector, 
\bea
\label{eq:Gamma_XYZ}
\nn
\mathcal{R}_{a}(f)&=&\mathcal{R}_{b}(f),\\
\Gamma_{ab}(f)&=&\Gamma_{cd}(f),
\eea
where $a,b,c,d={\rm X,Y,Z}$ with $a\neq b\neq c\neq d$. 

Based on the above derivation, the transfer function and \ac{ORF} of $\rm AET$ channel group 
\bea
\label{eq:Gamma_AET}
\nn
\mathcal{R}_{\rm A}(f)&=&\mathcal{R}_{\rm E}(f)=\mathcal{R}_{\rm X}(f)-\Gamma_{\rm XY}(f),\\
\nn
\mathcal{R}_{\rm T}(f)&=&\mathcal{R}_{\rm X}(f)+2\Gamma_{\rm XY}(f),\\
\Gamma_{\rm AE}(f)&=&\Gamma_{\rm AT}(f)=\Gamma_{\rm ET}(f)=0.
\eea
The last line of \eq{eq:Gamma_AET} shows that there is no correlation between the responses of $\rm A$, $\rm E$ and $\rm T$ to \ac{SGWB}, and
\be
\mathcal{R}_{a}(f)=-2\Gamma_{ab}(f),\quad f\ll f_{\ast}.
\ee
Thus, the first two lines of \eq{eq:Gamma_AET} reduce to
\bea
\label{eq:Gamma_XYZ_lf}
\nn
\mathcal{R}_{\rm A}(f)&=&\mathcal{R}_{\rm E}(f)=\frac{3}{2}\mathcal{R}_{\rm X}(f),\\
\mathcal{R}_{\rm T}(f)&=&o(\mathcal{R}_{\rm X}(f)),\quad f\ll f_{\ast},
\eea
which implies that $\rm T$ acts as a null channel. 

\begin{figure}[t]
	\centering
	\includegraphics[height=6cm]{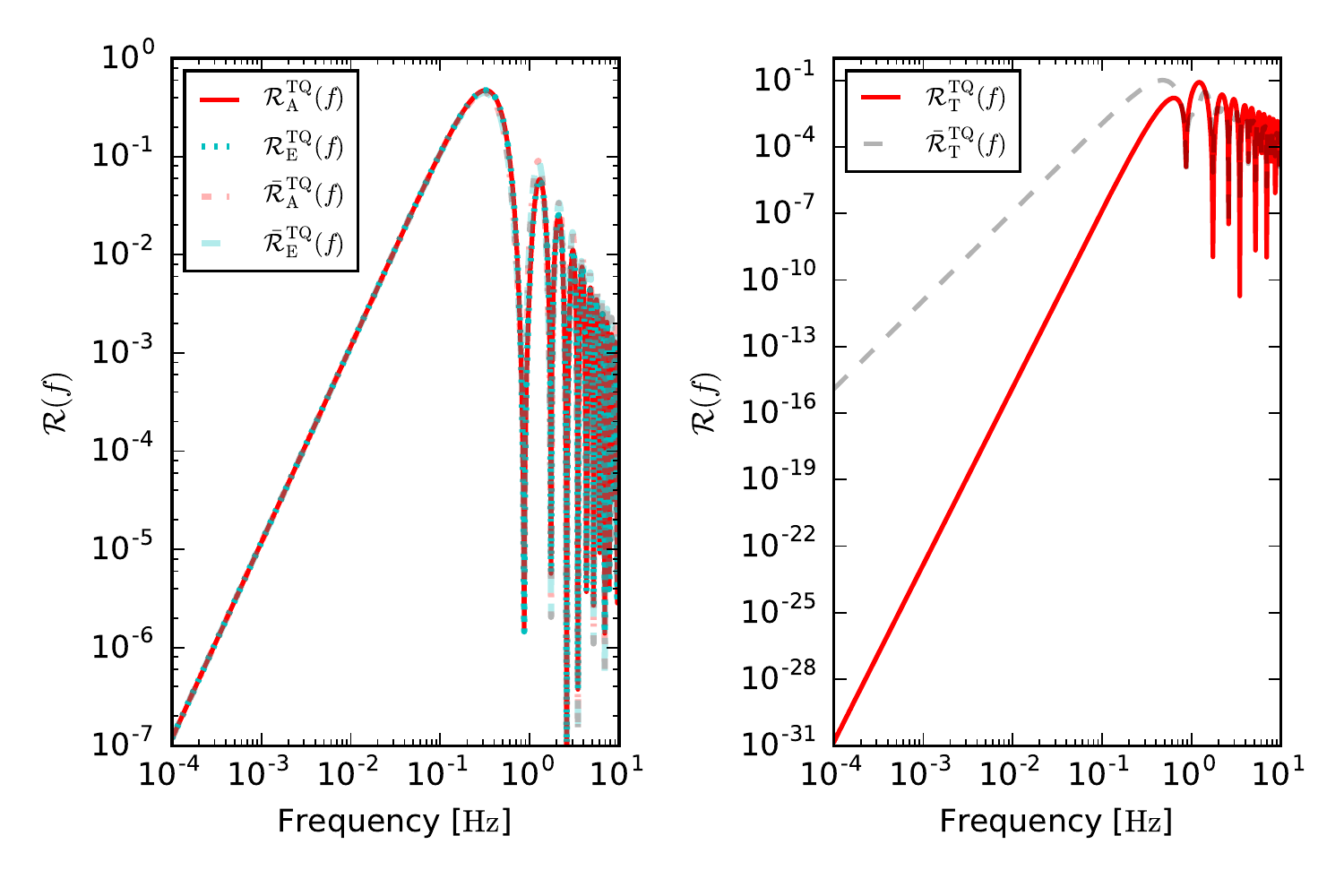}
	\caption{Transfer function of the channel group $\rm AET$ for TianQin.}
	\label{fig:ORF_TQ_AET}
\end{figure}

Furthermore, take TianQin as example, the transfer functions of channel group $\rm AET$ based on \eq{eq:F_aet} are shown in \fig{fig:ORF_TQ_AET}. 
And in order to illustrate the effect of reference interference site on the transfer function, we show the corresponding result within the dashed line, which employs the false response functions:
\bea
\label{eq:F_aet_false}
\nn
\bar{F}_{\rm A}^{P}(f,\hat{k},t_{0})&=&\frac{1}{\sqrt{2}}
\big[F_{\rm Z}^{P}(f,\hat{k},t_{0})-
F_{\rm X}^{P}(f,\hat{k},t_{0})\big],\\
\nn
\bar{F}_{\rm E}^{P}(f,\hat{k},t_{0})&=&\frac{1}{\sqrt{6}}
\big[F_{\rm X}^{P}(f,\hat{k},t_{0})-
2F_{\rm Y}^{P}(f,\hat{k},t_{0})
+F_{\rm Z}^{P}(f,\hat{k},t_{0})\big],\\
\bar{F}_{\rm T}^{P}(f,\hat{k},t_{0})&=&\frac{1}{\sqrt{3}}
\big[F_{\rm X}^{P}(f,\hat{k},t_{0})+
F_{\rm Y}^{P}(f,\hat{k},t_{0})
+F_{\rm Z}^{P}(f,\hat{k},t_{0})\big].
\eea
Under the low-frequency approximation, the misuse of analytical form has little effect on $\rm A/E$ channels, but a significant impact on $\rm T$ channel.  

\section{Channel noise level}\label{appen:noise}
For the one-way tracking built in the space-borne detector, the output mainly consists of four parts~\cite{1999ApJ...527..814A,Vallisneri:2004bn}:
\bea
\label{eq:dl_tot}
\nn
\delta l_{ij}(t)&=&\psi_{ij}(t)+C_{i}(t-L_{ij})-C_{j}(t)\\
&\quad&+n_{ij}^{\rm p}(t)+n_{ij}^{\rm a}(t-L_{ij})-n_{ji}^{\rm a}(t), 
\eea
where $L_{ij}$ is the armlength between satellite $i$ and $j$, $\psi_{ij}$ specifies the \ac{GW} signal. 
$C_{i}$, $n_{ij}^{\rm p}$ and $n_{ij}^{\rm a}$ are the noises caused by laser frequency, {\it aggregate} optical-path and {\it single} proof-mass acceleration, respectively~\cite{Krolak:2004xp}. 
Since the laser noise cancels in equal-arm Michelson and \ac{TDI} channels~\cite{Tinto:1999yr,1999ApJ...527..814A}, we will constrain our focus on other noises. 

Assuming that the \ac{PSD} of the same type of noise is equal~\cite{Chatterji:2006nh}:
\be
\langle \widetilde{n}^{\alpha}_{ij}(f)\widetilde{n}^{\beta * }_{kl}(f')\rangle=\frac{1}{2}\delta_{\alpha,\beta}\delta_{ij,kl}\delta(f-f')S_{\alpha}(f),
\ee
then the noise \ac{PSD} of the equal-arm Michelson
\be
\label{eq:P_m0}
\bar{P}_{\rm n_{\rm M}}(f)
=4S_{\rm p}(f)+8\bigg(\cos^{2}\big[\frac{f}{f_{\ast}}\big]+1\bigg)S_{\rm a}(f),
\ee
and for the channel group $\rm XYZ$~\cite{Vallisneri:2012np}
\bea
\label{eq:P_x}
\nn
\bar{P}_{\rm n_{\rm X,Y,Z}}(f)
&=&4\sin^{2}\big[\frac{f}{f_{\ast}}\big]\bar{P}_{\rm n_{\rm M}}(f)\\
&=&
16\sin^{2}\big[\frac{f}{f_{\ast}}\big]
\bigg[S_{\rm p}(f)+2\bigg(\cos^{2}\big[\frac{f}{f_{\ast}}\big]+1\bigg)S_{\rm a}(f)\bigg].
\eea

Furthermore, for channel group $\rm AET$~\cite{Vallisneri:2012np}:
\bea
\label{eq:P_aet}
\nn
\bar{P}_{\rm n_{\rm A,E}}(f)
\nn
&=&8\sin^{2}\big[\frac{f}{f_{\ast}}\big]
\bigg[\bigg(\cos\big[\frac{f}{f_{\ast}}\big]+2\bigg)S_{\rm p}(f)+2\bigg(\cos\big[\frac{2f}{f_{\ast}}\big]+2\cos\big[\frac{f}{f_{\ast}}\big]
+3\bigg)S_{\rm a}(f)\bigg],\\
\bar{P}_{\rm n_{T}}(f)
&=&32\sin^{2}\big[\frac{f}{f_{\ast}}\big]\sin^{2}\big[\frac{f}{2f_{\ast}}\big]
\bigg(S_{\rm p}(f)+4\sin^{2}\big[\frac{f}{2f_{\ast}}\big]S_{\rm a}(f)\bigg).
\eea
However, the two sides of the above equation (\eq{eq:P_m0}-\eq{eq:P_aet}) cannot be directly equal. 
Firstly, to make the component noises dimensionally consistent, the acceleration noise $S_{\rm dL}^{\rm a}(f)=S_{\rm a}(f)/(2\pi f)^{4}$~\cite{Babak:2021mhe}. 
Secondly, to make the \ac{GW} signals in \eq{eq:ht_2w} and \eq{eq:dl_tot} consistent, one should divide the optical-path noise and the acceleration noise by $2L$~\cite{Cornish:2001qi}. 
Following this rule, the {\it strain} noise \ac{PSD}
\bea
\label{eq:S_strain}
\nn
S_{\rm n}^{\rm p}(f)&=&\frac{S_{\rm p}(f)}{(2L)^{2}},\\
S_{\rm n}^{\rm a}(f)&=&\frac{S_{\rm dL}^{\rm a}(f)}{(2L)^{2}}=\frac{S_{\rm a}(f)}{(2L)^{2}(2\pi f)^{4}},
\eea
which are in units of $\rm Hz^{-1}$. Then we employ $S_{\rm n}^{\rm p}$ and $S_{\rm n}^{\rm a}$ instead of $S_{\rm p}$ and $S_{\rm a}$ to correct the strain noise \ac{PSD}:
\bea
\label{eq:Pn_all}
\nn
P_{\rm n_{\rm M}}(f)&=&\frac{1}{L^{2}}
\bigg[S_{\rm p}(f)+2\bigg(\cos^{2}\big[\frac{f}{f_{\ast}}\big]+1\bigg)
\frac{S_{\rm a}(f)}{(2\pi f)^{4}}\bigg],\\
P_{\rm n_{\rm X}}(f)
&=&
\nn
\frac{4\sin^{2}\big[\frac{f}{f_{\ast}}\big]}{L^{2}}
\bigg[S_{\rm p}(f)+2\bigg(\cos^{2}\big[\frac{f}{f_{\ast}}\big]+1\bigg)\frac{S_{\rm a}(f)}{(2\pi f)^{4}}\bigg],\\
\nn
P_{\rm n_{\rm A}}(f)
&=&\frac{2\sin^{2}\big[\frac{f}{f_{\ast}}\big]}{L^{2}}
\bigg[\bigg(\cos\big[\frac{f}{f_{\ast}}\big]+2\bigg)S_{\rm p}(f)+2\bigg(\cos\big[\frac{2f}{f_{\ast}}\big]+2\cos\big[\frac{f}{f_{\ast}}\big]
+3\bigg)\frac{S_{\rm a}(f)}{(2\pi f)^{4}}\bigg],\\
\nn
P_{\rm n_{T}}(f)
\nn
&=&\frac{8\sin^{2}\big[\frac{f}{f_{\ast}}\big]\sin^{2}\big[\frac{f}{2f_{\ast}}\big]}{L^{2}}
\bigg(S_{\rm p}(f)+4\sin^{2}\big[\frac{f}{2f_{\ast}}\big]\frac{S_{\rm a}(f)}{(2\pi f)^{4}}\bigg).
\eea
By setting $S^{\rm a}_{\rm tot}(f)=4S_{\rm a}(f)$ under the low-frequency approximation~\cite{Hu:2018yqb,Babak:2021mhe}, the noise \ac{PSD} of equal-arm Michelson is further expressed as
\be
P_{\rm n_{\rm M}}(f)=\frac{1}{L^{2}}
\bigg[S_{\rm p}(f)+
\frac{S^{\rm a}_{\rm tot}(f)}{(2\pi f)^{4}}\bigg],
\ee
where the total noise $S^{\rm a}_{\rm tot}(f)$ contains the noise at both ends of the link. 
\begin{figure}[t]
	\centering
	\includegraphics[height=6cm]{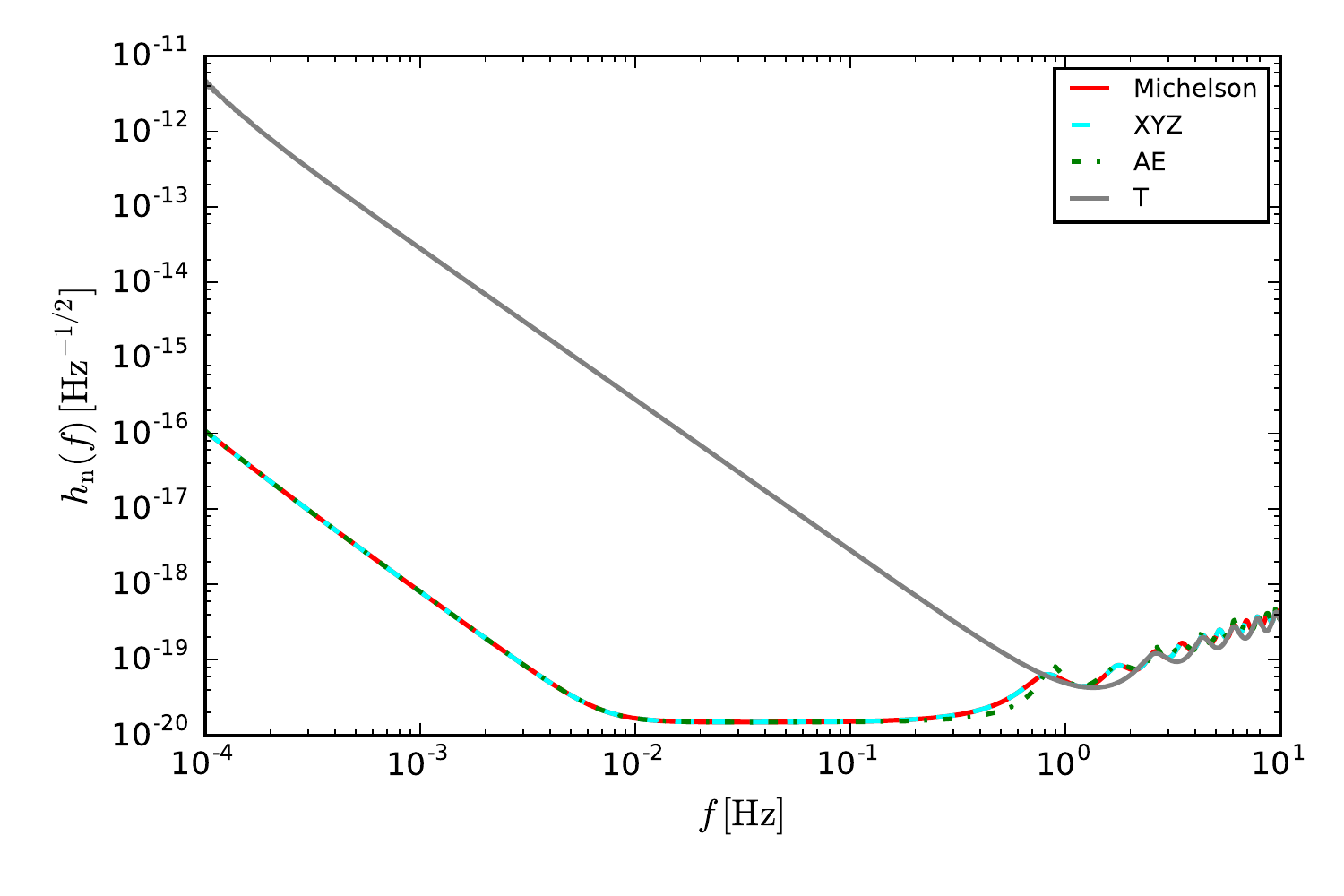}
	\caption{Sensitivity curve $h_{\rm n}(f)$ of the channel for TianQin.}
	\label{fig:Sensitivity_MXAT}
\end{figure}

On the other hand, the transfer functions of equal-arm Michelson and $\rm X$ can be converted by \eq{eq:F_X}:
\be
\label{eq:R_X2R_M}
\frac{\mathcal{R}_{\rm X}(f)}{\mathcal{R}_{\rm M}(f)}
=2\big(1-\cos\big[\frac{2f}{f_{\ast}}\big]\big)
=4\sin^{2}\big[\frac{f}{f_{\ast}}\big].
\ee
Combined with \eq{eq:S_nI}, \eq{eq:Gamma_XYZ_lf}, \eq{eq:Pn_all} and \eq{eq:R_X2R_M}, it can be inferred that the detection sensitivity of equal-arm Michelson, $\rm X$ and $\rm A$ are the same under the low-frequency approximation:
\be
S_{{\rm n}_{\rm M}}(f)=S_{{\rm n}_{\rm X}}(f)=S_{{\rm n}_{\rm A}}(f),\quad f\ll f_{\ast}.
\ee
In \fig{fig:Sensitivity_MXAT}, we show the sensitivity curves of the above channels for TianQin: $h_{\rm n}(f)=\sqrt{S_{{\rm n}}(f)}$. 
On one hand, the sensitivity curves of equal-arm and \ac{TDI} Michelson channels are the same below the characteristic frequency $f_{\ast}$. 
Although the sensitivity curve of $\rm T$ is much lower than that of other channels at low frequencies, both of them are proportional to $f^{-2}$. 
On the other hand, the sensitivity of Michelson channels drops around $3f_{\ast}$, beyond which $\rm T$ will be no longer treated as a noise monitor~\cite{Vallisneri:2007xa}. 

\section{Relevant derivation of detection method}\label{appen:Q_W}
For the output from noise-independent channel $I$ and $J$:
\be
\langle \widetilde{s}_{I}(f,t_{0})\widetilde{s}_{J}^{*}(f,t_{0})\rangle =\langle \widetilde{h}_{I}(f,t_{0})\widetilde{h}_{J}^{*}(f,t_{0})\rangle.
\ee
Note that, someone has discussed the subtraction of correlated noise~\cite{Thrane:2014yza,Coughlin:2016vor}. For simplicity, we prefer to neglect the correlated noise at this stage of the analysis. 
Then the expectation value and variance $\sigma^{2}(t)$ of the measurement can be calculated by
\bea
\label{eq:mu_IJ}
\nn
\mu(t_{0})&=&\langle S_{IJ}(t_{0})\rangle\\
\nn
&=&
\int_{t_{0}-T/2}^{t_{0}+T/2}{\rm d}t
\int_{-\infty}^{\infty}{\rm d}f\int_{-\infty}^{\infty}{\rm d}f'\,
\langle \widetilde{h}_{I}(f,t_{0})\widetilde{h}_{J}^{*}(f',t_{0})\rangle 
\widetilde{Q}_{IJ}(f',t_{0})e^{-{\rm i}2\pi(f-f')t_{0}}\\
\nn
&=&
\int_{t_{0}-T/2}^{t_{0}+T/2}{\rm d}t
\int_{0}^{\infty}{\rm d}f\int_{0}^{\infty}{\rm d}f'\,
\delta(f-f')P_{{\rm h}_{IJ}}(f,t_{0})\widetilde{Q}_{IJ}(f',t_{0})e^{-{\rm i}2\pi(f-f')t_{0}}\\
&=&T\int_{0}^{\infty}{\rm d}f\,
P_{{\rm h}_{IJ}}(f,t_{0})\widetilde{Q}_{IJ}(f,t_{0}),
\eea 
\bea
\label{eq:sigma_IJ}
\nn
\sigma^{2}(t_{0})
&=&
\langle S_{IJ}(t_{0})S_{IJ}(\eta_{0})\rangle-\langle S_{IJ}(t_{0})\rangle\langle S_{IJ}(\eta_{0})\rangle\\
\nn
&=&
\int_{t_{0}-T/2}^{t_{0}+T/2}{\rm d}t
\int_{\eta_{0}-T/2}^{\eta_{0}+T/2}{\rm d}\eta
\int_{-\infty}^{\infty}{\rm d}f\int_{-\infty}^{\infty}{\rm d}f'
\int_{-\infty}^{\infty}{\rm d}\omega\int_{-\infty}^{\infty}{\rm d}\omega'
\\
\nn
&&\times
\big[\langle\widetilde{s}_{I}(f,t_{0})\widetilde{s}_{I}^{*}(-\omega,\eta_{0})\rangle
\langle\widetilde{s}_{J}(-f',t_{0})\widetilde{s}_{J}^{*}(\omega',\eta_{0})\rangle
+\langle\widetilde{s}_{I}(f,t_{0})\widetilde{s}_{J}^{*}(\omega',\eta_{0})\rangle
\langle\widetilde{s}_{J}(-f',t_{0})\widetilde{s}_{I}^{*}(-\omega,\eta_{0})\rangle\big]\\
\nn
&&\times
\widetilde{Q}_{IJ}(f',t_{0})\widetilde{Q}_{IJ}^{*}(-\omega',\eta_{0})
e^{-{\rm i}2\pi(f-f')t}e^{-{\rm i}2\pi(\omega-\omega')\eta}\\
\nn
&=&
\frac{T}{2}\int_{0}^{\infty}{\rm d}f\,
\bigg[\big(P_{{\rm n}_{I}}(f)+P_{{\rm h}_{I}}(f,t_{0})\big)
\big(P_{{\rm n}_{J}}(f)+P_{{\rm h}_{J}}(f,t_{0})\big)+
|P_{{\rm h}_{IJ}}(f,t_{0})|^{2}\bigg]|\widetilde{Q}_{IJ}(f,t_{0})|^{2}\\
&=&\frac{T}{2}\int_{0}^{\infty}{\rm d}f\,
P_{{\rm n}_{I}}(f)P_{{\rm n}_{J}}(f)W_{IJ}(f,t_{0})|\widetilde{Q}_{IJ}(f,t_{0})|^{2},
\eea
where the correction function is shown in \eq{eq:W_IJ} and the channel noise is assumed to be stationary:
\be
\langle \widetilde{n}_{I}(f)\widetilde{n}^{*}_{J}(f')\rangle
=\frac{1}{2}\delta_{IJ}\delta(f-f')P_{{\rm n}_{I}}(f)
\ee

Combined with \eq{eq:mu_IJ} and \eq{eq:sigma_IJ}, the square of \ac{SNR} is obtained by
\be
\label{eq:snr_cc0}
\rho^{2}(t)=\frac{\mu^{2}(t)}{\sigma^{2}(t)}=
\frac{2\,T\big[\int_{0}^{\infty}{\rm d}f\,
	P_{{\rm h}_{IJ}}(f,t_{0})\widetilde{Q}_{IJ}(f,t_{0})\big]^{2}}
{\int_{0}^{\infty}{\rm d}f\,P_{{\rm n}_{I}}(f)
	P_{{\rm n}_{J}}(f)W_{IJ}(f,t_{0})|\widetilde{Q}_{IJ}(f,t_{0})|^{2}}.
\ee
By the definition of positive-definite inner product for any pair of complex function $A(f)$ and $B(f)$~\cite{Allen:1997ad}:
\be
(A,B):=\int_{0}^{\infty}{\rm d}f\,A(f)B^{*}(f)P_{{\rm n}_{I}}(f)P_{{\rm n}_{J}}(f)W_{IJ}(f,t_{0}),
\ee
\eq{eq:snr_cc0} turns to
\be
\rho^{2}(t)=2\,T
\frac{\big(\widetilde{Q}_{IJ}(f,t_{0}),\frac{P^{*}_{{\rm h}_{IJ}}(f,t_{0})}{P_{{\rm n}_{I}}(f)P_{{\rm n}_{J}}(f)W_{IJ}(f,t_{0})}\big)^{2}}{\big(\widetilde{Q}_{IJ}(f,t_{0}),\widetilde{Q}_{IJ}(f,t_{0})\big)}.
\ee
To maximize the \ac{SNR} yields the solution:
\be
\widetilde{Q}_{IJ}(f,t_{0})
=\lambda\frac{P^{*}_{{\rm h}_{IJ}}(f,t_{0})}{P_{{\rm n}_{I}}(f)P_{{\rm n}_{J}}(f)W_{IJ}(f,t_{0})}
\ee
with a real constant $\lambda$ and the correction function 
\bea
\label{eq:W_IJ0}
W_{IJ}(f,t_{0})
&=&1+\frac{P_{{\rm h}_{I}}(f,t_{0})P_{{\rm n}_{J}}(f)+P_{{\rm h}_{J}}(f,t_{0})P_{{\rm n}_{I}}(f)}{P_{{\rm n}_{I}}(f)P_{{\rm n}_{J}}(f)}+
\frac{P_{{\rm h}_{I}}(f,t_{0})P_{{\rm h}_{J}}(f,t_{0})+
|P_{{\rm h}_{IJ}}(f,t_{0})|^{2}}{P_{{\rm n}_{I}}(f)P_{{\rm n}_{J}}(f)}.
\eea
Then the optimal \ac{SNR} of cross-correlation method is given by \eq{eq:snr_cc}. 
Note that, the correlation time $T$ should be long enough that the \acp{PSD} of channel noise $P_{{\rm n}}(f)$ and \ac{SGWB} signal $P_{{\rm h}}(f,t_{0})$ are nearly invariable in the frequency region $\Delta f\sim1/T$. 

For a single TianQin-like detector, it needs to construct a specific correlator. 
We start with the auto-correlation of channel group $\rm AET$
\bea
\label{eq:ave_AET}
\nn
\langle s_{I}(f,t_{0})s_{I}(f',t_{0}) \rangle
&=&
\langle n_{I}(f)n_{I}(f')\rangle+\langle h_{I}(f,t_{0})h_{I}(f',t_{0})\rangle\\
\nn
&=&
\frac{1}{2}\delta(f-f')[P_{{\rm n}_{I}}(f)+P_{{\rm h}_{I}}(f,t_{0})],\\
\nn
\langle s_{\rm T}(f)s_{\rm T}(f') \rangle
&=&
\langle n_{\rm T}(f)n_{\rm T}(f')\rangle\\
&=&
\frac{1}{2}\delta(f-f')P_{{\rm n}_{\rm T}}(f),
\eea
where $I={\rm A,E}$. At first glance, the auto-correlation \ac{PSD} $P_{{\rm h}_{I}}$ can be obtained by subtracting the second row from the first row of \eq{eq:ave_AET}, which is under the assumption that $P_{{\rm n}_{I}}(f)=z_{I}(f)P_{\rm n_{T}}(f)$. 
Just follow that thought line, the reconstructed correlator for null-channel method is written as~\cite{Smith:2019wny}
\be
s_{0}(t,t')=s_{I}(t)s_{I}(t')-\langle n_{I}(t)n_{I}(t')\rangle.
\ee
The next step is to obtain the measurement:
\bea
\nn
K(t_{0})&=&\sum_{I=\rm A,E}\int_{t_{0}-T/2}^{t_{0}+T/2}{\rm d}t\int_{t_{0}-T/2}^{t_{0}+T/2}{\rm d}t'
s_{0}(t,t')Q_{II}(t-t')\\
&\approx&\sum_{I=\rm A,E}
\int_{t_{0}-T/2}^{t_{0}+T/2}{\rm d}t
\int_{-\infty}^{\infty}{\rm d}f\int_{-\infty}^{\infty}{\rm d}f'
\,[\widetilde{s}_{I}(f,t_{0})\widetilde{s}_{I}^{*}(f',t_{0})-\langle\widetilde{n}_{I}(f,t_{0})\widetilde{n}_{I}^{*}(f',t_{0})\rangle]
\widetilde{Q}_{II}(f',t_{0})e^{-{\rm i}2\pi(f-f')t_{0}},
\eea
of which the expectation value $\mu(t)$ and the variance $\sigma^{2}$ are
\bea
\nn
\mu(t_{0})&:=&\langle K(t_{0})\rangle\\
\nn
&=&\sum_{I=\rm A,E}\int_{t_{0}-T/2}^{t_{0}+T/2}{\rm d}t\int_{-\infty}^{\infty}{\rm d}f
\int_{-\infty}^{\infty}{\rm d}f'
\,\langle\widetilde{h}_{I}(f,t_{0})\widetilde{h}^{*}_{I}(f',t_{0})\rangle \widetilde{Q}_{II}(f',t_{0})
e^{-{\rm i}2\pi (f-f')t_{0}} \\
&=&2\,T\int_{0}^{\infty}{\rm d}f\,P_{{\rm h}_{I}}(f,t_{0})\widetilde{Q}_{II}(f,t_{0}),\\
\nn
\sigma^{2}(t_{0})&=&
\langle K(t_{0})K(\eta_{0})\rangle-\langle K(t_{0})\rangle\langle K(\eta_{0})\rangle\\
\nn
&=&\sum_{I=\rm A,E}
\int_{t_{0}-T/2}^{t_{0}+T/2}{\rm d}t
\int_{\eta_{0}-T/2}^{\eta_{0}+T/2}{\rm d}\eta
\int_{-\infty}^{\infty}{\rm d}f\int_{-\infty}^{\infty}{\rm d}f'
\int_{-\infty}^{\infty}{\rm d}\omega\int_{-\infty}^{\infty}{\rm d}\omega'
\\
\nn
&&\times
\bigg[\left\langle\big(\widetilde{s}_{I}(f,t_{0})\widetilde{s}_{I}^{*}(f',t_{0})
-\langle\widetilde{n}_{I}(f,t_{0})\widetilde{n}_{I}^{*}(f',t_{0})\rangle\big)
\big(\widetilde{s}_{I}(\omega,\eta_{0})\widetilde{s}_{I}^{*}(\omega',\eta_{0})
-\langle\widetilde{n}_{I}(\omega,\eta_{0})\widetilde{n}_{I}^{*}(\omega',\eta_{0})\rangle\big)
\right\rangle\\
\nn
&&-\langle \widetilde{h}_{I}(f,t_{0})\widetilde{h}_{I}^{*}(f',t_{0}) \rangle
\langle \widetilde{h}_{I}(\omega,\eta_{0})\widetilde{h}_{I}^{*}(\omega',\eta_{0}) \rangle\bigg]
\widetilde{Q}_{II}(f',t_{0})\widetilde{Q}^{*}_{II}(\omega',\eta_{0})
e^{-{\rm i}2\pi(f-f')t}e^{-{\rm i}2\pi(\omega-\omega')\eta}\\
\nn
&=&
T\int_{0}^{\infty}{\rm d}f\,
2\big(P_{{\rm n}_{I}}(f)+P_{{\rm h}_{I}}(f,t_{0})\big)^{2}|\widetilde{Q}(f,t_{0})|^{2}\\
&=&2\,T\int_{0}^{\infty}{\rm d}f\,
P^{2}_{{\rm n}_{I}}(f)W_{I}(f,t_{0})|\widetilde{Q}_{II}(f,t_{0})|^{2},
\eea
where the correction function
\bea
W_{I}(f,t_{0})
&=&\bigg(1+\frac{P_{{\rm h}_{I}}(f,t_{0})}{P_{{\rm n}_{I}}(f)}\bigg)^{2}.
\eea
Compared with \eq{eq:W_IJ0}, the correction function $W_{I}(f,t_{0})$ does not involve the cross-term.

When the filter function
\be
\widetilde{Q}_{II}(f,t_{0})
=\lambda\frac{P_{{\rm h}_{I}}(f,t_{0})}{P^{2}_{{\rm n}_{I}}(f)W_{I}(f,t_{0})}
\ee
with a real constant $\lambda$, the optimal \ac{SNR} of null-channel method is obtained.
%----------------------------------------------------%%

\ew
%%%%%%%%%%%%%%%%%%%%%%%%%%%%%%%%%%%%%%%%%%%%%%%%%%%%%%%%%%%%%%%%
%%%% ²Î¿¼ÎÄÏ×¿ªÊ¼ %%%%%%%%%%%%%%%%%%%%%%%%%%%%%%%%%%%%%%%%%%%%%%
%%%%%%%%%%%%%%%%%%%%%%%%%%%%%%%%%%%%%%%%%%%%%%%%%%%%%%%%%%%%%%%%
\normalem
\bibliographystyle{apsrev4-1}
%%%%%%%%%%%%%%%%%%%%%%%%%%%%%%%%%%%%%%%%%%%%%%%%%%%%%%%%%%%%%%%%
%%% ½«²Î¿¼ÎÄÏ×Ìí¼Óµ½ reference.bib ÎÄ¼þÀï£¬ÔÚÕâÀïµ÷ÓÃ %%%%%%
%%%%%%%%%%%%%%%%%%%%%%%%%%%%%%%%%%%%%%%%%%%%%%%%%%%%%%%%%%%%%%%%
\bibliography{TQnote}

%%%%%%%%%%%%%%%%%%%%%%%%%%%%%%%%%%%%%%%%%%%%%%%%%%%%%%%%%%%%%%%%
%%%% ²Î¿¼ÎÄÏ×½áÊø %%%%%%%%%%%%%%%%%%%%%%%%%%%%%%%%%%%%%%%%%%%%%%
%%%%%%%%%%%%%%%%%%%%%%%%%%%%%%%%%%%%%%%%%%%%%%%%%%%%%%%%%%%%%%%%
\end{document}